\documentclass[a4paper,onecolumn,10pt,accepted=2025-07-09]{quantumarticle}
\pdfoutput=1
\usepackage[utf8]{inputenc}
\usepackage[english]{babel}
\usepackage[T1]{fontenc}
\usepackage{amsmath}
\usepackage{hyperref}
\usepackage{mathrsfs}
\usepackage{bm}
\usepackage{cancel}
\usepackage{dsfont}
\usepackage{physics}
\usepackage{caption}
\usepackage{subcaption} 
\usepackage{tikz}
\usepackage{lipsum}
\usepackage[normalem]{ulem}
\usepackage{times,amsmath,amsfonts,amssymb,latexsym,amsthm,bbm}
\usepackage{graphicx,epsf}
\usepackage[toc,page,header]{appendix}
\usepackage{minitoc}
\usepackage[numbers]{natbib}
\usepackage{color}
\usepackage{url}
\newcommand{\be}{\begin{equation}}
\newcommand{\ee}{\end{equation}}
\newcommand{\ba}{\begin{eqnarray}}
\newcommand{\ea}{\end{eqnarray}}
\newcommand{\ignore}[1]{}

\newcommand{\PSTAB}{\mathsf{PSTAB}}

\newcommand{\pur}{\operatorname{Pur}}

\newcommand{\exv}{\mathbb{E}}
\newcommand{\hi}{\mathcal{H}}
\newcommand{\id}{\mathbbm{1}}
\newcommand{\ot}{\otimes}
\newcommand{\pauli}{\mathbb{P}}
\newcommand{\stab}{{\rm STAB}}
\newcommand{\cl}{\mathcal{C}(d)}
\newcommand{\mlin}{M_{\rm lin}}
\newcommand{\elin}{E_{\rm lin}}
\newcommand{\perm}{{\rm T}}
\newcommand{\slambda}{\mathscr{S}_{\bm{\lambda}}}
\newcommand{\psil}{\ket{\psi^{\bm{\lambda}}}}

\newcommand{\lin}{\rm lin}

\def\norm#1{\Vert #1\Vert}
\def\CC{{\rm\kern.24em \vrule width.04em height1.46ex depth-.07ex
    \kern-.29em C}}
\def\P{{\rm I\kern-.25em P}}
\def\RR{{\rm
         \vrule width.04em height1.58ex depth-.0ex
         \kern-.04em R}}

\def\bbbc{{\mathchoice {\setbox0=\hbox{$\displaystyle\rm C$}\hbox{\hbox
to0pt{\kern0.4\wd0\vrule height0.9\ht0\hss}\box0}}
{\setbox0=\hbox{$\textstyle\rm C$}\hbox{\hbox
to0pt{\kern0.4\wd0\vrule height0.9\ht0\hss}\box0}}
{\setbox0=\hbox{$\scriptstyle\rm C$}\hbox{\hbox
to0pt{\kern0.4\wd0\vrule height0.9\ht0\hss}\box0}}
{\setbox0=\hbox{$\scriptscriptstyle\rm C$}\hbox{\hbox
to0pt{\kern0.4\wd0\vrule height0.9\ht0\hss}\box0}}}}
\def\bbbz{{\mathchoice {\hbox{$\sf\textstyle Z\kern-0.4em Z$}}
{\hbox{$\sf\textstyle Z\kern-0.4em Z$}}
{\hbox{$\sf\scriptstyle Z\kern-0.3em Z$}}
{\hbox{$\sf\scriptscriptstyle Z\kern-0.2em Z$}}}}

\newtheorem{prop}{Proposition}

\begin{document}
\setcounter{secnumdepth}{3}


\title{Entanglement and Stabilizer entropies
of random bipartite pure quantum states} 
\author[1,2]{Daniele Iannotti}
\author[1,2]{Gianluca Esposito}
\author[3,4]{Lorenzo Campos Venuti}
\author[1,2,3]{Alioscia Hamma}

\affil[1]{Scuola Superiore Meridionale, Largo S. Marcellino 10, 80138 Napoli, Italy}
\affil[2]{Istituto Nazionale di Fisica Nucleare (INFN) Sezione di Napoli}
\affil[3]{Università degli Studi di Napoli Federico II , Dipartimento di Fisica Ettore Pancini}
\affil[4]{Department of Physics and Astronomy, University of Southern California, Los Angeles, USA}

\begin{abstract}
The interplay between non-stabilizerness and entanglement in random states is a very rich arena of study for the understanding of quantum advantage and  complexity. In this work, we tackle the problem of such interplay in random pure quantum states. We show that while there is a strong dependence between entanglement and magic, they are, surprisingly, perfectly uncorrelated. We compute the expectation value of non-stabilizerness given the Schmidt spectrum (and thus entanglement). At a first approximation, entanglement determines the average magic on the Schmidt orbit. However, there is a finer structure in the average magic distinguishing different orbits where the flatness of entanglement spectrum is involved.

\end{abstract}
\maketitle

\doparttoc
\faketableofcontents

\section{Introduction}
Entanglement has long been regarded a cornerstone of quantum information science, distinguishing quantum mechanics from classical theories and serving as a pivotal resource for quantum technologies \cite{horodecki2009QuantumEntanglement}. 
Since the advent of the stabilizer formalism, it has been clear that entanglement is not enough to provide computational advantage \cite{gottesman1998heisenberg}.
Such a formalism identifies a set of states, called stabilizer states, which have the peculiar feature of being efficiently simulable using classical computational resources despite being possibly highly entangled \cite{gottesman1998heisenberg, nielsen_chuang_2010}. States away from the set of stabilizer states are a fundamental resource for universal quantum computation. 
Indeed, the distance from the set of the stabilizer states \cite{chitambar2019QuantumResourceTheories} defines the non-stabilizer resource, which also plays an important role in characterizing the complexity of quantum states and processes \cite{campbell2011CatalysisActivationMagic,bravyi2016TradingClassicalQuantum,beverland2020LowerBoundsNonClifford,leone2022stabilizer,leone2023nonstabilizerness,goto2021ChaosMagica,garcia2023resource, PhysRevLett.123.020401}.

Entanglement and magic are thus distinct but interrelated resources for understanding the structure and behavior of quantum states.
Previous investigations into this interplay have yielded several key insights. Notably, entanglement can be computed exactly for stabilizer states \cite{fattal2004entanglement, hamma2005BipartiteEntanglementEntropic}, establishing a foundational link between entanglement and classical simulability.
The probability distribution of entanglement in random stabilizer states  \cite{Dahlsten:2006ojk} establishes another connection between entanglement and the free stabilizer resources. A series of works show that this kind of entanglement has a simple pattern \cite{chamon2014EmergentIrreversibilityEntanglement} and that entanglement complexity arises when enough non-stabilizer resources (also known as \emph{magic}) are injected  {\cite{ chamon2014EmergentIrreversibilityEntanglement, LeoOliZhouHam21, ZhouYangHammaChamon2020, lami2025quantum,gu_2024_pseudomagic}}, also by measurement in monitored quantum circuits \cite{OLIVIERO2021127721, PRXQuantum.5.030332}.
Furthermore, a connection between non-stabilizerness and the entanglement response of quantum systems, i.e. anti-flatness of the reduced density operator, has been identified \cite{tirrito2024quantifying}, highlighting how these resources interact under system dynamics. Additionally, a computational phase separation has been observed, categorizing quantum states into two distinct regimes: entanglement-dominated and magic-dominated phases \cite{PRXQuantum.6.020324}.

This work aims at establishing some exact results in the interplay between entanglement and non-stabilizerness in random pure quantum states. We will utilize Stabilizer Entropy (SE) \cite{leone2022stabilizer} as the unique computable monotone for non-stabilizerness \cite{Leone_Bittel_2024}. We start with a simple consideration that has, though, profound consequences: the separable state with maximal SE is much less resourceful than the average pure state drawn from the Haar measure. 

Haar-random states are typically highly entangled so we see that most entangled states possess a (much) higher SE than the maximum-SE separable state. This means that entanglement makes room for non-stabilizerness and that the two resources must feature a rich interplay.

\begin{figure}
    \centering
    \includegraphics[width=0.8\linewidth]{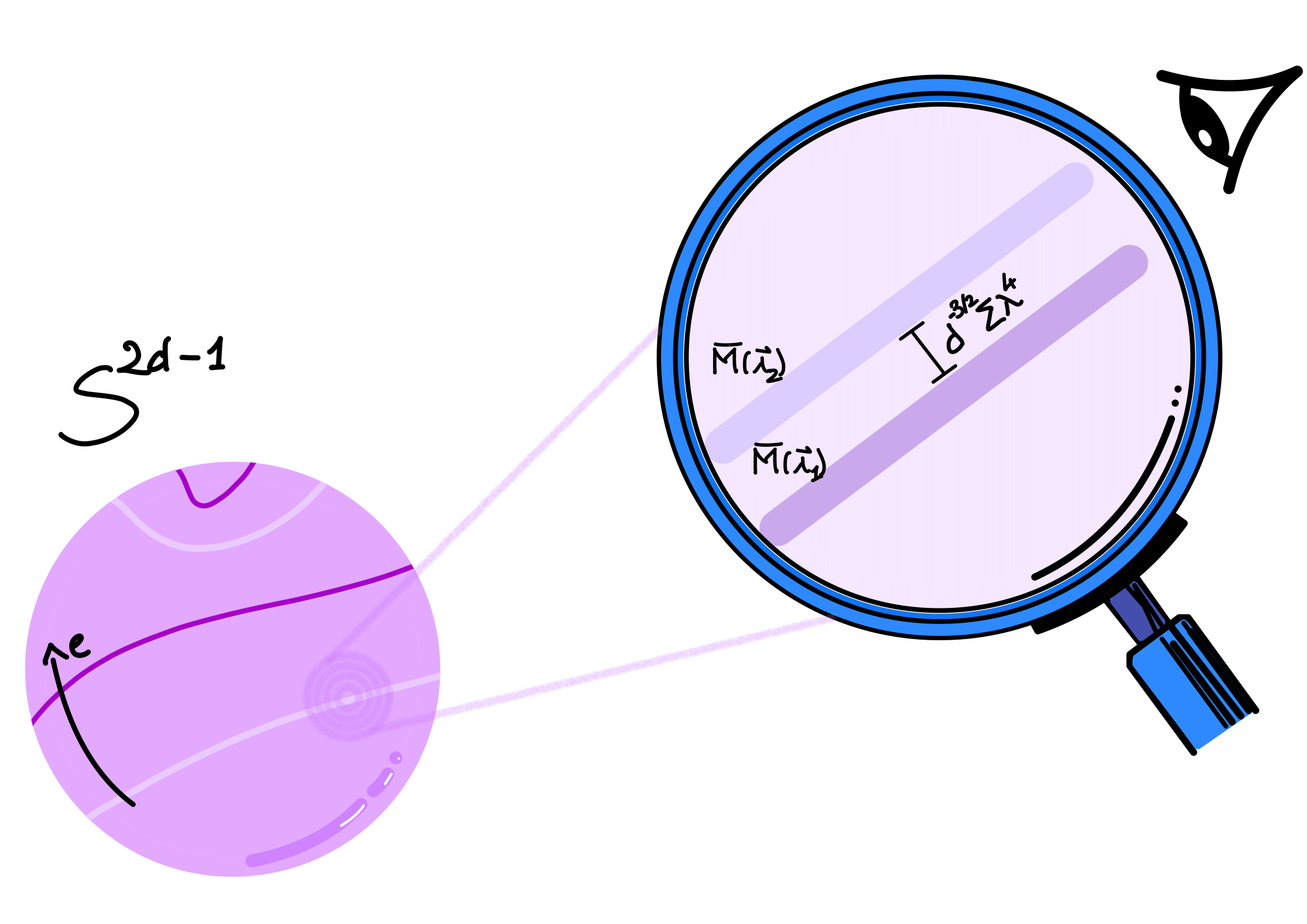}
    \caption{Illustration of Schmidt orbits (colored pink and violet lines on the sphere) categorized by their entanglement in a coarse-grained view (left) and the fine structure detailed by the averaged magic evaluated over each orbit (right).}
    \label{fig:graphic}
\end{figure}

In this paper, we give a quantitative and analytical analysis of the dependence of these two quantities in random pure states. A numerical analysis with similar scope was recently presented in \cite{Szombathy:2025euv}.
The main result of this work is the surprising fact that magic and entanglement are exactly uncorrelated (in their linear versions), yet dependent, and a way to picture this intricate dependence is via the foliation of the Hilbert space of bipartite pure states using the concept of Schmidt orbits, see Fig.~\ref{fig:graphic}.

The paper is structured as follows: In Section \ref{ch:haar} we set up the notation, the necessary tools such as Haar measure and a survey of known results on linear entanglement and SE.  
Section \ref{ch:corr} discusses the surprising results that the linear entanglement and SE have null covariance.
In Section \ref{ch:schmidt} we address the dependence between entanglement and SE presenting a calculation of the average SE over Schmidt orbits and present results about its typically. 
In Section \ref{example} we compute the averaged magic at fixed entanglement for the specific case of a $d=2\times d_B$-dimensional bipartite Hilbert space.
Finally, Section \ref{ch:flat} discusses the broader picture in which the result of the previous section stands in the literature, namely the connection with anti-flatness.


\section{Setting the stage}
\label{ch:haar}
In this section, we give a description of the tools and notation used in the paper. 
We consider a finite dimensional Hilbert space $\mathcal{H}=\mathcal{H}_A \otimes \mathcal{H}_{B}$. Concretely {,} $\mathcal{H}$ is a collection of $n$ qubits, i.e.~$d=\dim{\mathcal{H}}=2^n$ and $d_{A,B}=\dim{\mathcal{H}_{A,B}}=2^{n_{A,B}}$. Owing to normalization, the set of pure states on  $\mathcal{H}$ can be identified with the hypersphere $S^{2d-1}\subset \mathbb{R}^{2d}$, more precisely {,} any pure state can be identified with a Hopf circle on the sphere $S^{2d-1}$. On this manifold (isomorphic to $\mathbb{C}P^{d-1}$) there is a unique, unitarily invariant measure $d\psi$. This measure is induced by the Haar measure $dU$ on the corresponding unitary group $U(d)$ when applied to a fiducial state $|\psi_0\rangle \in \mathcal{H}$, i.e.
\begin{equation}
\int d\psi f(|\psi\rangle ) = \int_{U(d)} dU f(U|\psi_0\rangle ) \, , \label{eq:Haar0}
\end{equation}
for any integrable function $f$. A function from the set of pure states to $\mathbb{R}$ becomes a real random variable when $\mathbb{C}P^{d-1}$ is equipped with this uniform measure. Expectation values of $f$ are computed with Eq.~\eqref{eq:Haar0} via $\mathbb{E}_\psi [f(|\psi\rangle )]=\mathbb{E}_U [f(U|\psi_0\rangle )] =\int dU f(U|\psi_0\rangle )$. 

\subsection{Entanglement and Linear Entanglement Entropy}
In this paper we consider two particular functions on the set of pure states quantifying, respectively, entanglement and non-stabilizerness. 
In order to be more symmetric with respect to our choice of magic measure (see below), we consider the linear entanglement entropy $E_\mathrm{lin}$ as a measure of entanglement  {for a pure state $\psi$.} $\elin$  is defined as 
\begin{equation}
    \elin(\psi)=1-\pur(\psi_A)=1-\Tr_A[(\Tr_B \psi)^2]
    \label{eq:elin}
\end{equation}
where $\psi_A=\Tr_B \psi$.  {Linear entanglement (and linear stabilizer entropies) are important from the theoretical, computational and experimental point of view. For the former, questions of typicality (see also the following) are addressed by analyzing the statistical moments of such resources in the Hilbert space \cite{popescu2006EntanglementFoundationsStatistical, leone2021QuantumChaosQuantum}; in the same way the resource inducing capabilities of quantum operations can be computed through their linear versions through the resource power. As an example, the entangling power of a unitary is defined as the average entanglement induced on the set of factorized states \cite{PhysRevA.62.030301, PhysRevA.69.062319},  non-stabilizer power as the average SE on the set of stabilizer states \cite{leone2022stabilizer} and coherence power averaging on the set of incoherent states \cite{PhysRevA.95.052306}. Mathematically, linear entropy is the only entanglement measure that has the same functional form when expressed in terms of the Wigner function \cite{PhysRevE.62.4665}. From the computational point of view, they have the advantage of being polynomials in the state and can be computed more efficiently, especially in fermionic systems \cite{PhysRevA.75.032301}. Finally, and very importantly, only linear quantities are accessible experimentally as only they can be expressed as expectation values of Hermitian (and thus, linear) operators \cite{PhysRevLett.108.110503,elben2018RenyiEntropiesRandom,islam2015MeasuringEntanglementEntropy, oliviero2022MeasuringMagicQuantum} without requiring full state tomography. 
}
The linear entanglement entropy is also related to the \textit{concurrence} via $\mathcal{C}_A(\psi)=\sqrt{2\elin(\psi)}$ \cite{bengtsson_geometry_2006}, and to the \textit{entanglement 2-R\'enyi entropy}, via $S_2(\psi_A)=-\log(1-\elin(\psi))$.

In order to define the measure of magic that we are going to consider we briefly review some basic facts about the stabilizer formalism. 

\subsection {Non-stabilizerness and Stabilizer Entropy}

The \textit{Pauli group} on $n$ qubits is defined as

\begin{equation}
    \Tilde{\pauli}_n:=\{\pm 1, \pm i\}\times\{\id, X, Y, Z\}^{\ot n}\,.
\end{equation}
 In this work, single tensor product of Pauli matrices is referred to as \textit{Pauli strings} of \textit{Pauli operators} $\pauli_n$. Pauli strings play an important role, since they also provide an orthogonal basis for the space of linear operators on $\hi$.
 Let us denote the Clifford group as the normalizer of the Pauli group, namely

 \begin{equation}
     \mathcal{C}(d):=\{C\in U(d)\,|\, CPC^\dag=P'\in \pauli(d)\,,\forall P\in \pauli(d)\}\,.
 \end{equation}

Given these elements, the set of \textit{stabilizer states} of $\hi$ is defined as the orbit of the Clifford group through the computational basis states  $\ket{i}$ (namely, the eigenstates of the $Z$ operator): in formulas,
\begin{equation}
    \stab:=\{C\ket{i}\,,\,C\in \cl\}\,.
\end{equation}
Equivalently, a pure stabilizer state is defined as the common eigenstate of $d$ mutually commuting Pauli strings. 

Stabilizer states share some properties about the computational complexity of simulating quantum processes using classical resources. These properties are summarized by the \textit{Gottesman-Knill theorem}, which states that any quantum process that can be represented with initial stabilizer states upon which one performs (i) Clifford unitaries, (ii) measurements of Pauli operators, (iii) Clifford operations conditioned on classical randomness, can be perfectly simulated by a classical computer in polynomial time \cite{gottesman1998heisenberg}. 
Since the set of stabilizer states is by definition closed under Clifford operations, some resources, such as unitary operations outside the Clifford group or states not in $\stab$, need to be injected in the quantum system in order to make it  universal. These non-stabilizer resources, referred to as \textit{non-stabilizerness} (or \textit{magic}) of the state, have been proven to be a useful resource for universal quantum computation \cite{bravyi2005UniversalQuantumComputation} and  several measures have been proposed to quantify it \cite{howard2017ApplicationResourceTheory, veitch2014ResourceTheoryStabilizer}.

 {In this work, we focus on the unique
computable monotone for non-stabilizerness, namely the  \textit{Stabilizer Entropy} (SE) - in particular, the 2-Stabilizer R\'enyi Entropy  $M_2$ \cite{leone2022stabilizer} - and its linear counterpart $\mlin$. The linearized SE $\mlin$ has all the advantages discussed above for linear entropies, for instance in the definition of 
non-stabilizing power  \cite{leone2022stabilizer}, the establishment of typical behavior, or its experimental measurement \cite{oliviero2022MeasuringMagicQuantum}.
 Moreover, $\mlin$ is also known to be significant in the process of creation of $\varepsilon$-approximate state $t$-designs \cite{vairogs2025extracting_randomness}.}

Starting from the probability distribution $\Xi_P(\ket{\psi}):=\frac{1}{d} {\Tr}^2(P\ketbra{\psi})$, with $P\in\pauli_n$, associated to the tomography of the quantum state $\psi$, the $2$-Stabilizer R\'enyi Entropy for pure states is defined as 
\begin{equation}
    M_2(\psi) :=-\log\left[d\sum_{P\in \pauli_n} \Xi_P(\psi)^2\right]=-\log\left[\frac{1}{d}\sum_{P} \Tr^4(P\psi)\right]=-\log\left[d\Tr(Q\psi^{\otimes 4})\right]\,,
\end{equation}
with $Q:=\frac{1}{d^2}\sum_P P^{\ot 4}$, whereas the linear SE is defined as follows:
\begin{equation}
    \mlin(\psi):=1-d\Tr(Q\psi^{\ot 4})\equiv 1 -{\rm SP}(\psi) \, , 
\end{equation}
where $\mathrm{SP}(\psi)$ is the stabilizer purity. Both of these measures are: (i) faithful, i.e. $M_2(\psi)=\mlin(\psi)=0 \Leftrightarrow \psi \in \stab$; (ii) non-increasing over free operations, that is the operations preserving $\stab$; (iii) additive (or multiplicative for the stabilizer purity SP) under tensor product, namely $M_2(\psi\otimes\sigma)=M_2(\psi)+M_2(\sigma)$. Both the linear and the logarithmic SEs are regarded as good monotones for the pure-state resource theory of stabilizer computation \cite{Leone_Bittel_2024}. 

\subsection{Known results about Entanglement and Non-stabilizerness}
\label{sub:known_results}
Now that we defined our measures of entanglement and magic we review some known facts about them when considered independently. 
The first two moments of $\elin$ and $\mlin$ have been computed. In particular one has \cite{lubkin1978entropy,page1993AverageEntropySubsystem,lloyd1988BlackHolesDemons} with $(\psi_U:= U\psi U^\dagger$)
\begin{equation}
    \exv_U[\elin(\psi_U)]=1-\frac{d_A+d_B}{d_A d_B+1}.
\end{equation}
Since in the regime $d_A \gg 1$ and $d_B\gg 1$, $\exv_U[\elin(\psi_U)]\approx1$, this indicates that on average, except for the case when one of the two dimensions is small, generic states in the Hilbert space are nearly maximally entangled. 

The variance of $\elin$ has been computed in \cite{scott2003entangling} and reads
\begin{equation}
    \Delta^2 \elin  = \exv_U \left [\elin^2(\psi_U)\right ] -  \exv_U ^2\left [\elin(\psi_U)\right ]
    = \frac{2 \left(d_A^2-1\right) \left(d_B^2-1\right)}{(d+1)^2 (d+2) (d+3)} = O\Big(\frac{1}{d^2}\Big)\,.
\end{equation}

Since the variance tends to zero as $d\to \infty$, the distribution of $\elin$ becomes increasingly peaked as the dimension grows. Using Chebyshev's inequality one can show that the probability that $\elin$ is very different from its average value is very small. One says then that $\elin$ \emph{typically} is close to its average or simply that the random variable is \emph{typical}.  

Another way of proving the typicality of $\elin$ is by use of L\'evy's lemma. 
In Lemma III.8 of \cite{hayden2006AspectsGenericEntanglement} the authors show that the Lipschitz constant of $\sqrt{\Tr(\psi_A^2)}$ is upper bounded by $2$. Using a similar reasoning or with little modification 
one can show that the same holds for $\Tr(\psi_A^2)$ with Lipschitz constant $\eta \leq 2$, hence one can state that entanglement also exhibits the stronger typicality offered by L\'evy's lemma.
    
Average and variance (corresponding to the first two cumulants) are also known for the linear magic $\mlin$ {, which can be used to study magic spreading in random quantum circuits \cite{turkeshi2025magic}}. In this case one has (see \cite{leone2022stabilizer,zhu2016CliffordGroupFails})

\begin{equation}
\begin{split}
   \exv_{U}[\mlin(\psi_U)]&= 1-d\,\mathbb{E}_{U}[\Tr(Q\psi_{U}^{\otimes4})]
= 1-\frac{4}{d+3}= O(1)\,,\\
\Delta^{2}\mlin&=  \frac{96(d-1)}{(d+3)^{2}(d+5)(d+6)(d+7)}=O\Big(\frac{1}{d^{4}}\Big)\,, 
\end{split}
\end{equation}

for details on how to obtain the variance the reader is referred to Appendix \ref{var_calc} or to \cite{zhu2016CliffordGroupFails}. 
 {Indeed, apart from the first two moments, a characterization of the Pauli spectrum was carried out for physically relevant states in \cite{PhysRevB.111.054301}.}

Since also the variance of the magic vanishes when $d\to \infty$ (and with a greater exponent than that of the entanglement), using Chebyshev's inequality one can prove typicality of $\mlin$. Alternatively, using the Lipshitz constant found  in \cite{zhu2016CliffordGroupFails} and  Levy's lemma, one can easily obtain   that \cite{leone2022stabilizer}

\begin{equation}
     \operatorname{Pr}(|\mlin(\ket\psi)-\exv[\mlin]| \geq \epsilon) \leq 3 \exp \left(-\frac{ \epsilon^2 d}{729\pi}\right)\,.
\end{equation}


\section {Covariance between Entanglement and Magic}
\label{ch:corr}
The strongest motivation behind this work is to further understand  the relationship between magic and entanglement in random states. 
The need for clarification of the interplay between magic and entanglement has been fueled by various numerical evidence regarding the maximum achievable value of magic through separable states.  The (pure) single-qubit state achieving maximum SE is the so-called \textit{golden state}, defined as:
\begin{equation}
    \ketbra{G} :=\frac{1}{2}\left(\id+\frac{X+Y+Z}{\sqrt{3}}\right)\,.
\end{equation}
We can see then that the separable state with maximal SE is much less resourceful than the average state in the Hilbert space. Since one can show that $\exv_U[{M}_2(\psi_U)]  \ge \log(d+3) - \log 4 $ one has the following bound
\begin{equation}
        M_2(\ketbra{G}^{\otimes n})=n \log{\frac{3}{2}} < {\log\frac{d+3}{4}} \le  \exv_U[{M}_2(\psi_U)]\,,
\end{equation}
and a similar situation holds for $\mlin$, namely 
\begin{equation}
 \mlin(\ketbra{G}^{\otimes n})= 1- \left(\frac{2}{3} \right)^n \le 1 - \frac{4}{d+3}\,.
\end{equation}
We see that as most states in the Hilbert space are very entangled, entanglement is a precondition to have high - magic states. In other words, this shows that there is a strong interplay between non-stabilizerness and entanglement.  

One would be tempted to say, in fact, that $\mlin$ and $\elin$ are strongly correlated. We therefore set out to compute the covariance
\begin{equation}
    \begin{split}
        {\rm Cov}(\elin,\mlin)&:=\exv_U[\elin(\psi_U)\mlin(\psi_U)]- \exv_U[\elin(\psi_U)] \exv_U[\mlin(\psi_U)]
        \\&= d\, \exv_U \left[ \Tr[(Q\otimes \perm_2 ^{A}\otimes \id_B ^{\ot 2})\psi_U ^{\ot 6}]\right] -\frac{4 (d_A+d_B)}{(d+1) (d+3)}\,.
    \end{split}
\end{equation}
Surprisingly, we find the following result.
\begin{prop}
The linear entanglement and linear magic, when seen as random variables over the states of pure states equipped with the uniform measure, are uncorrelated. Namely:
    \begin{equation}\label{corr0}
     {\rm Cov}(\elin,\mlin)=0\, .
\end{equation}
\end{prop}
In order to prove the above proposition we divided all the 720 permutations of the symmetric group of six elements $S_6$ into conjugacy classes and performed traces over all of them (see Appendix \ref{cov_calc} for details).
The same result of lack of covariance does not hold for the non-linearized versions of the variables. However, there is numerical evidence by analysis of Fig.~\ref{fig:EM_all} and the numerical analysis on the fluctuations of magic and entanglement in the recent \cite{Szombathy:2025euv} that suggest that the covariance of the logarithmic versions of entanglement and SE tends to zero in the thermodynamic limit.

\begin{figure}[h!]
    \centering
    \includegraphics[width=6cm]{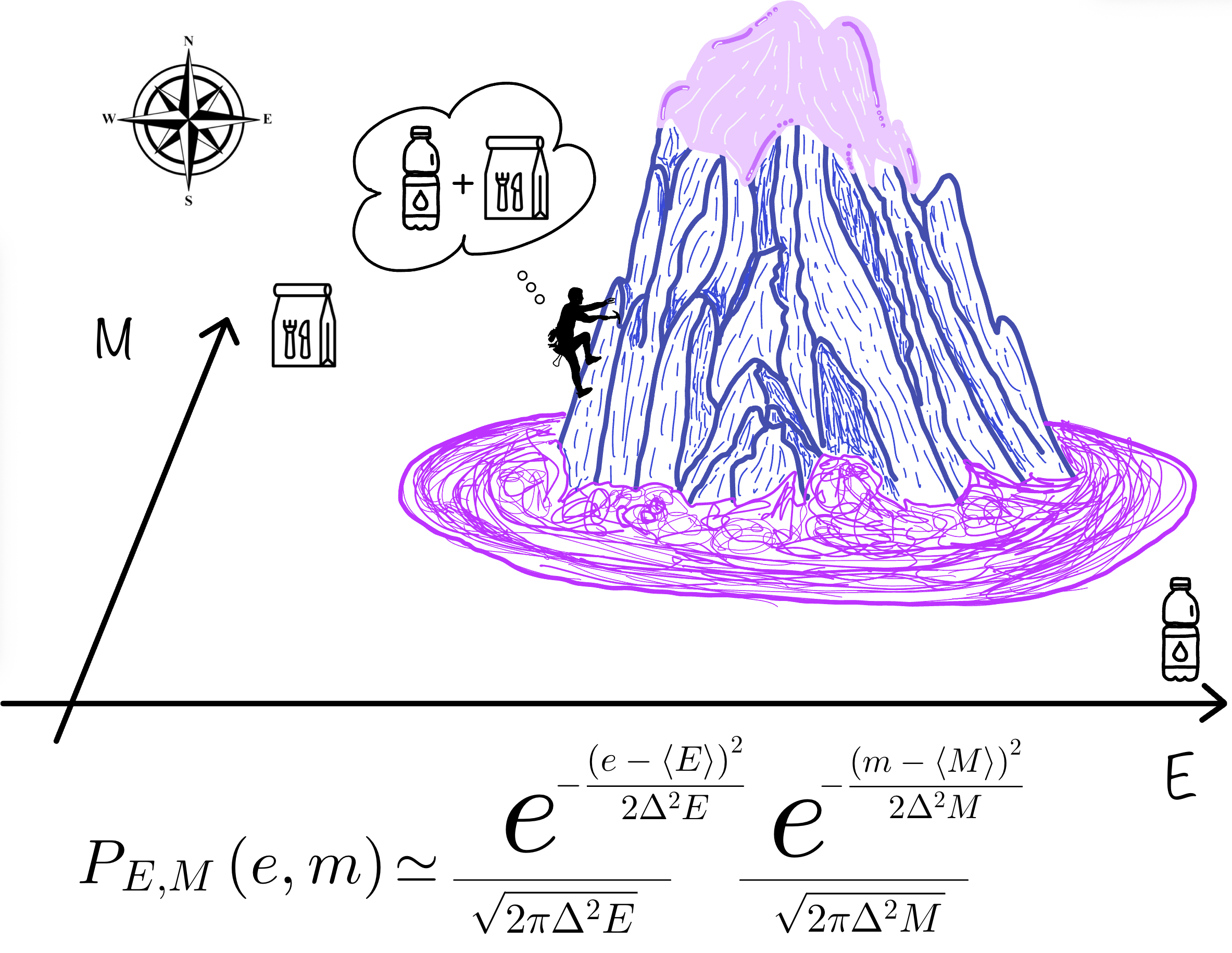}
    \includegraphics[width=8cm]{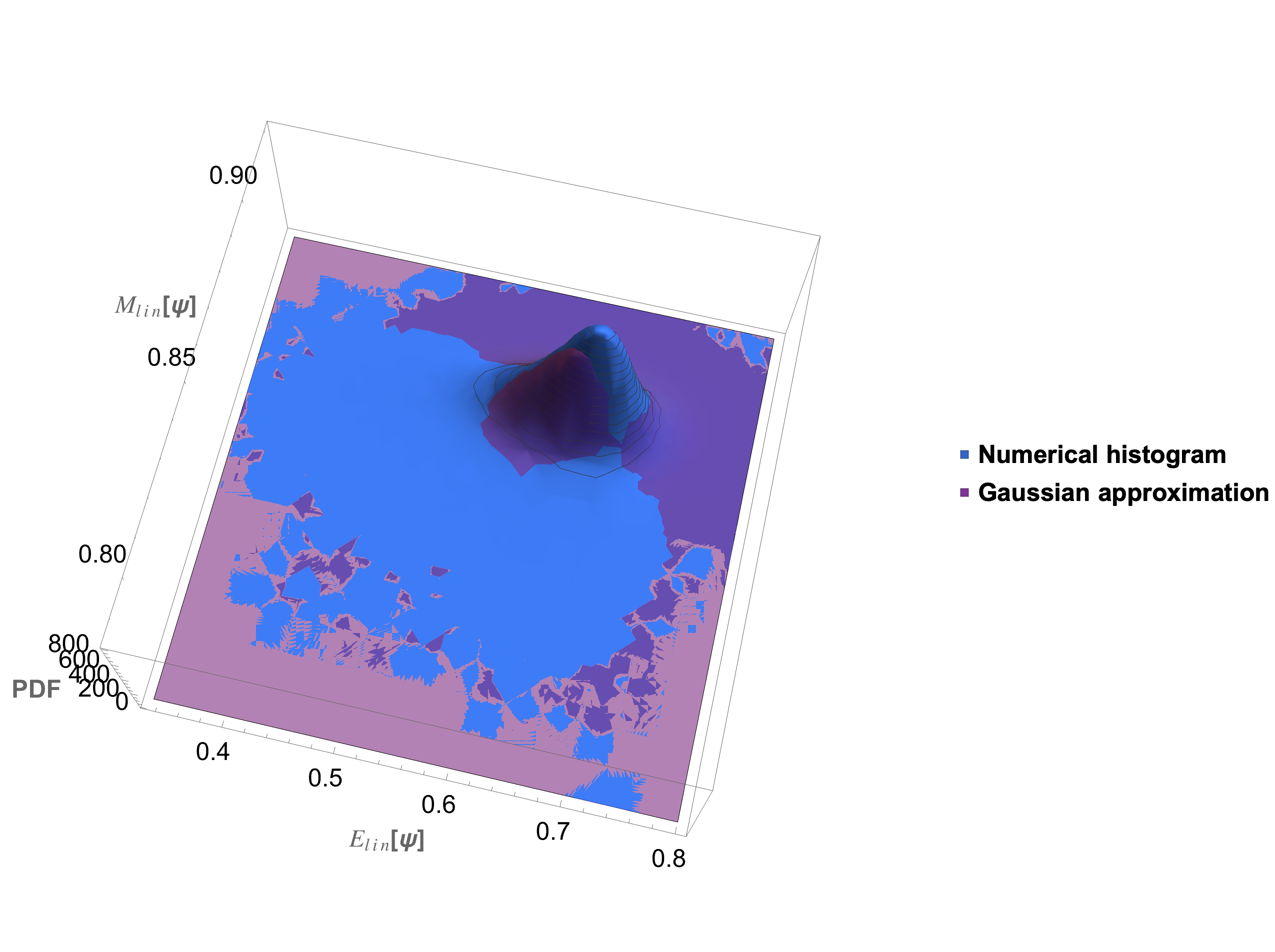}
\caption{ {\textit{Left panel}. Pictorial representation of the approximated joint probability distribution for the Haar measure of magic ($M$) and entanglement ($E$) in the large $d$ limit since the correlation is identically zero. The principal axes of the ellipse are aligned with the axes of entanglement and magic, as no linear correlation exists between the two variables. \textit{Right panel}. Joint probability density function of $\mlin$ and $\elin$ obtained with $N_{\rm sample}=10^6$ Haar random pure states $\ket{\psi}$ for 5 qubits with (12|345) bipartition (blue), and the gaussian approximation Eq.~\eqref{eq:gauss_approx} (purple) for the same probability density function. Note that the level curves of the PDF are approximately ellipses with axes parallel to the $\elin$, $\mlin$ axes.} }
    \label{fig:mountain}
\end{figure}

This fact, although unexpected, does not imply that entanglement and magic are independent as notably independence implies that the variables are uncorrelated but not vice-versa\footnote{For a toy example of this phenomenon, consider a standard Gaussian random variable $X$ and its square $Y=X^2$ (obviously dependent on $X$). The covariance between $X$ and $Y$ is equal to the third moment, which is zero due to the symmetry of the Gaussian distribution.}. In fact, the covariance detects only the \emph{linear} dependence between the variables.
 {As we saw in Section~\ref{sub:known_results}, both entanglement and magic become (separately) concentrated in the limit of large dimensionality. This does not necessarily imply that the correlation between the two variables tends to zero. In fact, even though two random variables, $X$ and $Y$, are highly concentrated for large dimensionality, i.e.~$P_X(x) \to  \delta(x)$ and $P_Y(y) \to  \delta(y)$, in the limit $d\to\infty$, their correlation,  $\text{Corr}(X,Y)=\frac{\text{Cov}(X,Y)}{\Delta X \Delta Y}\to C\neq0$, may be finite in the same limit. See Appendix \ref{App:example_correlation} for a specific example.
In terms of the joint characteristic function $\chi_{\elin,\mlin}(\boldsymbol{\xi}) := \exv_U [ e^{ i (\xi_1 \elin+\xi_2 \mlin)}]$ what we have shown is that $\chi_{\elin,\mlin} (\boldsymbol{\xi}) = \exp[- \boldsymbol{\xi}^T C \boldsymbol{\xi}/2 +R(\boldsymbol{\xi})]$, where the correlation matrix $C$ is diagonal and $R(\boldsymbol{\xi})$ depends on joint moments of order three or greater. Under the assumption that $R$ is small we can write $e^{R(\boldsymbol{\xi})}=1+R(\boldsymbol{\xi})+\ldots $ and evaluate the Fourier transform of $\chi_{\elin,\mlin}(\boldsymbol{\xi})$ perturbatively. We arrive at the following result:
\begin{equation}
P_{\elin,\mlin}(e,m) \simeq \frac{1}{2\pi \Delta E \Delta M} \exp[-\frac{(e-\langle \elin \rangle)^2}{2\Delta^2 E} - \frac{(m-\langle \mlin \rangle)^2}{2\Delta^2 M} ] \, ,   \label{eq:gauss_approx}  
\end{equation}
where, with the symbol $\simeq$, we mean that the right-hand side gives the correct moments up to second order, i.e.~for quantities of the form $\exv_U [\elin^p \mlin^q]$ with $p,q,$ integers satisfying $ 0 \le p+q\le2 $. Note that the Gaussian approximation, Eq.~\eqref{eq:gauss_approx}, cannot be correct for all the moments when either $d_A$ or $d_B$ is small, as an explicit calculation (see Eq.~\eqref{eq:PE2_marginal} below) shows that the marginal is not Gaussian in this case. However, we checked that the Gaussian approximation is accurate in the limit of large dimensionality (see Appendix \ref{app:gauss_approx}) and already for $d_A=2^2,d_B=2^3$ reproduces the numerical data quite precisely (see Fig.~\ref{fig:mountain} right panel). }
{The operational meaning of Eq.~\eqref{eq:gauss_approx}, is that the level curves $P_{\elin,\mlin}(e,m) = \mathrm{const.}$, up to corrections of third order moments or greater, are ellipses centered around $\langle \elin \rangle, \, \langle \mlin \rangle$ with semi-axes parallel to the $\elin, \, \mlin$ axes. This means that if we want to increase the probability density we should aim at the maximum of $P_{\elin,\mlin}(e,m)$ (the peak of the mountain in Fig.~\ref{fig:mountain}) in steps of either constant entanglement, or constant magic, pretty much how a mountaineer would do to reach a peak, trying to climb as much as possible either horizontally or via the path of maximum steepness in order to avoid false incline where balancing is trickier\footnote{Clearly there can be situations in the mountain where climbing a false incline is the best strategy.} (see Fig.~\ref{fig:mountain} left panel for a pictorial representation). 
In other words, out of all the possible measures of entanglement and magic, $\elin$, and $\mlin$ are "intrinsically orthogonal". }

To obtain a better understanding of the dependence between entanglement and magic, a more detailed study is required. This will be the content of the next section.

\section {Magic conditioned by entanglement and averages over Schmidt orbits}
\label{ch:schmidt}

After considering the previous results about the absence of covariance between entanglement and magic, we would like to characterize their dependence. 
In this section, we examine the average behavior of magic given the non-local properties of the state. For this goal, we fix the Schmidt coefficients in a bipartition and compute the average magic on the Schmidt orbit. In this way, we find some technical results on non-local magic, and thus the average interplay between magic and entanglement.

Two random variables are independent if the realization
of one does not affect the probability distribution of the other.
The most complete way to quantify the dependence of two random variables
is to compute their conditional probabilities, such as $P\left(\mlin=m|\elin=e\right)\equiv P(m|e)$,
the probability distribution of $\mlin$ given that entanglement has the value $e$. 
The variables are independent if $P(m|e)= P(m)$ for all values of $e$. In the following, we set out to obtain a partial information regarding this conditional probability, namely its average. 

Consider a pure state $|\psi\rangle$ in the bipartite system $\mathcal{H}=\mathcal{H}_{A}\otimes\mathcal{H}_{B}$.
Fixing a product basis in $\mathcal{H}$, $|\phi_{i}\rangle_{A}\otimes|\chi_{j}\rangle_{B}$
the pure state is represented by a rectangular matrix $\Psi_{i,j}=\langle \phi_{i},\chi_{j}|\psi\rangle$.
We now perform a singular value decomposition on this matrix, $\Psi=U_{A}DU_{B}^{T}$,
where $U_{A,B}$ are unitaries in $\mathcal{H}_{A,B}$ and $D$ is
the rectangular matrix with the singular values, $\boldsymbol{\lambda}$,
of $\Psi$ on the diagonal. Assuming $d_{B}\ge d_{A}$, this corresponds
to the Schmidt decomposition 
\begin{equation} \ket{\psi}=\sum_{i=1}^{d_{A}}\sqrt{\lambda_{i}}\,U_{A}\otimes U_{B}\ket{\phi_{i}}\otimes \ket{\chi_{i}}\,.
\end{equation}
If $|\psi\rangle$ is sampled uniformly with the Haar measure, i.e.~$|\psi\rangle=U_{AB}\,|\phi_1,\chi_1\rangle$
with $U_{AB}$ Haar distributed, what is the measure induced by the Schmidt decomposition? It turns out that it is a product measure \citep{bengtsson_geometry_2006}
\begin{equation}
P(\boldsymbol{F}^{A,B})\times P\left(\boldsymbol{\lambda}\right)\,,\label{eq:product_measure}
\end{equation}
where the first factor denotes the natural, unitarily invariant distribution on the flag manifold $\boldsymbol{F}^{A,B}=U(d_{A})\otimes U(d_{B})/U(1)^{d_{A}}$, while the probability distribution $P(\boldsymbol{\lambda})$ on the simplex
$\Delta_{d_{A}-1}$ has been computed in \citep{lloyd_complexity_1988,zyczkowski_induced_2001,giraud2007distribution,giraud2007purity}.
The quotient with $U(1)^{d_{A}}$ arises because the
Schmidt decomposition is invariant under the gauge symmetry $U_{A}\mapsto U_{A}V,$
$U_{B}\mapsto U_{B}{V'}^*$ where $V$ is a $d_{A}\times d_{A}$ diagonal
unitary matrix with $d_{A}$ phases while $V'$ is $d_{B}\times d_{B}$
has the same phases on the diagonal and zero on the remaining entries.

For the functions we are interested in, functions of $M_{\mathrm{lin}}(|\psi\rangle)$ and $E_{\mathrm{lin}}(|\psi\rangle)$, 
Haar integration over the flag manifold $\boldsymbol{F}^{A,B}$ coincides
with Haar integration over $U(d_{A})$ and $ U(d_{B})$ (see Appendix \ref{sec:Haar}). Then we have the following result
\begin{align}
P\left(m| e\right) & :=\frac{\int_{U(d_{A}d_{B})}dU_{AB} \, \delta\left(M_{\mathrm{lin}}(U_{AB}|\psi\rangle)-m\right)\,\delta\left(E_{\mathrm{lin}}(U_{AB}|\psi\rangle)-e\right)}{P(e)}\nonumber \\
 & =\frac{1}{P(e)}\int_{U(d_{A})}dU_{A}\int_{U(d_{B})}dU_{B}\int_{\Delta_{d_{A}-1}}d\boldsymbol{\lambda}\,P(\boldsymbol{\lambda})\nonumber \\
 & \times\delta\left(M_{\mathrm{lin}}(U_{A}\otimes U_{B}|\psi^{\bm{\lambda}}\rangle)-m\right)\delta\left(1-\sum_{i=1}^{d_{A}}\lambda_{i}^{2}-e\right)\, {,}\label{eq:prob_cond}
\end{align}
where $|\psi^{\bm{\lambda}}\rangle=\sum_{i=1}^{d_{A}}\sqrt{\lambda_{i}}|\phi_{i},\chi_{i}\rangle$
is a reference state with given Schmidt coefficients. Moreover, since
$E_{\mathrm{lin}}$ depends only on the Schmidt coefficients and not
on the unitaries $U_{A},U_{B}$, we have
\begin{equation}
P(e)=\int_{\Delta_{d_{A}-1}}d\boldsymbol{\lambda}\,P(\boldsymbol{\lambda})\,\delta\left(1-\sum_{k=1}^{d_{A}}\lambda_{k}^{2}-e\right).
\end{equation}
Looking at Eq.~\eqref{eq:prob_cond}, we note that we reduced the
task to that of computing averages over the \emph{Schmidt orbits} 
\begin{equation}
    \mathscr{S}_{\boldsymbol{\lambda}}:=\{U_{A}\otimes U_{B}|\psi^{\bm{\lambda}}\rangle:U_{A}\in U(d_{A}),U_{B}\in U(d_{B})\}\,.
\end{equation}
This is reminiscent of the approach in \cite{biswas_fidelity_2024}.
As anticipated, we will content ourselves with the first moment
of the above conditional probability, namely 
\begin{equation}
\begin{split}
\Tilde{M}(e) & :=\int dm\,m\,P\left(m|e\right)\\
 & =\frac{1}{P(e)}\int_{U(d_{A})}dU_{A}\int_{U(d_{B})}dU_{B}\int_{\Delta_{d_{A}-1}}d\boldsymbol{\lambda}\,P(\boldsymbol{\lambda})\\
 & \times M_{\mathrm{lin}}(U_{A}\otimes U_{B}|\psi^{\bm{\lambda}}\rangle)\,\delta\left(1-\sum_{k=1}^{d_{A}}\lambda_{k}^{2}-e\right)\,.
 \end{split}
\end{equation}


Define now the average magic over Schmidt orbits, $\bar{M}(\boldsymbol{\lambda})$, as 
\begin{equation}\label{avgM}
\bar{M}(\boldsymbol{\lambda}):=\mathbb{E}_{U_{A},U_{B}}\left[M_{\mathrm{lin}}(U_{A}\otimes U_{B}|\psi^{\bm{\lambda}}\rangle)\right]\,,
\end{equation}
then the average magic at given entanglement $e$ is given by 
\begin{equation}
\label{eq:m_e}
\Tilde{M}(e)=\frac{1}{P(e)}\int_{\Delta_{d_{A}-1}}d\boldsymbol{\lambda}\,P(\boldsymbol{\lambda})\,\bar{M}(\boldsymbol{\lambda})\, \delta\left(1-\sum_{k=1}^{d_{A}}\lambda_{k}^{2}-e\right)\,.
\end{equation}


{It is interesting that the average magic given a certain entanglement value depends crucially on the function $\Bar{M}(\boldsymbol{\lambda})$. This function has the operational meaning of being the amount of magic that can be created on the state $|\psi\rangle$ (with Schmidt coefficient $\boldsymbol{\lambda}$), on average over the Schmidt orbits. 
It is customary in resource theory to define the power of a unitary as the  resource that can be created  \emph{on average} by the map. This is because in absence of perfect control, the average resource can be a much more useful quantity compared to the maximum or minimum resource. 
For example, in \cite{leone2022stabilizer}, the (non)-stabilizing power of a unitary operator is defined as the average SE that can be created acting with the unitary on the set of stabilizer states:
\begin{align}
\mathcal{M}\left(U\right) & :=\frac{1}{\left|\PSTAB\right|}\sum_{|\psi\rangle\in\PSTAB}\mlin\left(U|\psi\rangle\right) \nonumber \\
& = \mathbb{E}_{|\psi\rangle \in \PSTAB}\left[\mlin(U |\psi\rangle )\right ].
\label{eq:stab_power}    
\end{align}
}

 {Since here we are interested in the average magic and at fixed entanglement resources, we consider the resource monotone $\mlin$  on the set of free  operations for the other resource, entanglement, that is factorized unitaries of the form $U_A \otimes U_B$. This is essentially the dual of Eq.~\eqref{eq:stab_power}:
\begin{equation}
 \mathbb{E}_{U_{A},U_{B}}\left[M_{\rm lin}(U_{A}\otimes U_{B}|\psi\rangle)\right] 
=\mathbb{E}_{U_{A},U_{B}}\left[M_{\rm lin}(U_{A}\otimes U_{B}|\psi^{\boldsymbol{\lambda}}\rangle)\right]=:\bar{M}(\boldsymbol{\lambda})  ,   
\end{equation}
where we used the fact that the result of the average depends only on the Schmidt coefficients of $|\psi\rangle$. 
Thus $\bar{M}(\boldsymbol{\lambda})$ captures the non-local character of non-stabilizer resources: the amount of magic that can be injected on a state by local unitary operations on average. As we will show below in Sec.~\ref{sec:typicality} this is the \emph{typical} value of linear magic created by factorized operations $U_A \otimes U_B$ on the state $|\psi^{\boldsymbol{\lambda}}\rangle$, i.e.~the variable $\mlin(U_A \otimes U_B |\psi^{\boldsymbol{\lambda}}\rangle)$ \emph{concentrates}.  
It is obviously  bounded from above by the maximum SE that can be unitarily injected locally in a state and from below by the residual SE after erasure by local unitaries: 
\ba
 m^{{\rm NL}} (\boldsymbol{\lambda}):= \min_{U_A\otimes U_B} m(U_A\otimes U_B\ket{\psi}) \le \bar{M}(\boldsymbol{\lambda}) \le M^{{\rm NL}} (\boldsymbol{\lambda}):= \max_{U_A\otimes U_B} m(U_A\otimes U_B\ket{\psi})\,,
\ea
for any non-stabilizer monotone $m$. 
Importantly, $m^{{\rm NL}}  (\boldsymbol{\lambda})$ characterizes the reduced density spectrum flatness, the capacity of entanglement, and is a measure of back-reaction in the AdS/CFT correspondence \cite{cao2024gravitational}.}
Then we have the following proposition (see Appendix  \ref{SE_mean_app} for a proof).

\begin{prop}
    Given a bipartite $n=n_A+n_B$-qubit  Hilbert space $\mathcal{H}=\mathcal{H}_A \otimes \mathcal{H}_B$ with dimensions $d_A \equiv 2^{n_A} \leq 2^{n_B}\equiv d _B$,   the average value of $M_{\lin}$ over the Schmidt orbit $U_A\ot U_B$ on the reference state $|\psi^{\bm{\lambda}}\rangle$ reads 
    \begin{equation}\label{eq:m_UAUB}
        \begin{split}
            \Bar{M}({\bm \lambda})&= 1-d_A d_B\,\exv_{U_A,U_B}[\Tr(Q(U_A\ot U_B)^{\ot 4} \psi^{{\bm \lambda}\ot 4} (U_A\ot U_B)^{\dag \ot 4})]\\&=\alpha+\beta e+\gamma e^2+\delta\sum_i \lambda_i ^3+\mu\sum_i \lambda_i ^4\,,
        \end{split}
    \end{equation}
    where
    \begin{equation}\label{coeff}
        \begin{split}
            \alpha&=1-\frac{4}{3} \left(\frac{8}{d_A  d_B  }+\frac{5}{(d_A +3) (d_B  +3)}-\frac{1}{(d_A -3) (d_B  -3)}\right)\,,\\
            \beta&=\frac{8}{(d_A +3) (d_B +3)}+\frac{16}{d_A  d_B }\,,\\
            \gamma&=-\left(\frac{4 (d_A  d_B +9)}{\left(d_A ^2-9\right) \left(d_B^2-9\right)}+\frac{8}{d_A  d_B }\right)\,,\\
            \delta&=-\frac{96 \left(d_A ^2+d_A  d_B +d_B ^2-9\right)}{d_A  \left(d_A ^2-9\right) d_B  \left(d_B ^2-9\right)}\,,\\
            \mu&=\frac{24 (d_A +d_B )}{\left(d_A ^2-9\right) \left(d_B ^2-9\right)}\,,
        \end{split}
    \end{equation}
 and the normalization implies $\sum_i \lambda_i =1$ while the linear entanglement entropy is $e=1-\sum_i \lambda_i ^2$.
    \label{prop:M_AB}
\end{prop}
Notice that if  $\ket{\psi}$ is separable $M({\bm \lambda})$ reads
\begin{equation}
    \Bar{M}({\bm{\lambda}}=(1,0,\ldots,0))
    =1-\frac{16}{(d_A+3) (d_B+3)}\,,
\end{equation}
whereas for maximally entangled states one has
\begin{equation}
    \Bar{M}\left({\bm{\lambda}}=\left(\frac{1}{d_A},\ldots,\frac{1}{d_A}\right)\right)=\frac{d_A^3 d_B^3-9 d_A^3 d_B-4 d_A^2 d_B^2+24 d_A^2+12 d_A d_B-24}{d_A^3 (d_B-3) d_B (d_B+3)}\, .
\end{equation}
In the large d limit with $d_A=d_B=\sqrt{d}$, Eq.~\eqref{eq:m_UAUB} becomes 
    \begin{equation}\label{mlambda_asymptotic}
        \begin{split}
         \bar{M}({\bm{\lambda}})&=_{d \gg 1} 1-\frac{12 e^2-24 e+16}{d}-\frac{48(e-1)}{d^{3/2}}-\frac{36\left(3 e^2-6 e+4\right)}{d^2}-\frac{864(e-1)}{d^{5/2}}\\&-\frac{48}{d^{3/2}}\left[\frac{6}{\sqrt{d}} \sum_i \lambda_i^3-\sum_i\lambda_i^4\right]+O\left(\frac{1}{d^3}\right)\,.
    \end{split}
    \end{equation}
As one can see, in the large $d$ limit, the leading term (up to $O(1/d)$) depends only on the entanglement of the orbit. It seems that at first glance, the average magic of states with a fixed Schmidt decomposition can be approximated by a quadratic polynomial of its entanglement: however, there are dependences on higher moments of the Schmidt distribution that distinguish the orbits. 
In the region where $d_B \gg d_A \gg 1 $ the asymptotic expansion becomes
\begin{equation}
\begin{split}
   \bar{M}({\bm{\lambda}})&=_{d_B \gg d_A \gg 1}  1 -\frac{12 e^2-24 e+16}{d}-\frac{24 (e-1)}{d d_A}-\frac{12 \left(3 e^2-6 e+4\right)}{d d_A^2}\\
   &-\frac{24}{d_A d}\left[\frac{4}{d_A}\sum_i \lambda_i ^3-\sum_i \lambda_i ^4\right] + O\left( \frac{1}{d^2 d_A ^2} \right)\,,
   \end{split}
\end{equation}
hence, even in this case the behaviour is similar to that of Eq.~\eqref{mlambda_asymptotic}.

 {Notice that when $|\psi^{\bm{\lambda}}\rangle$ is a stabilizer state, all $\alpha$-Rényi entropies of $\psi_A$ are equal to $E\equiv E(|\psi^{\bm{\lambda}}\rangle)= -\Tr(\psi_A \log \psi_A)=-\log(1-e)$, which has an integer value \cite{fattal2004entanglement}.
Hence, in this case $e=1-2^{-E}$, and Eq.~\eqref{eq:m_UAUB} is directly a function of the (non-linear) entanglement $E$.
}

\begin{figure}[h!]
    \centering
    \begin{subfigure}[b]{0.48\textwidth}
        \centering
        \includegraphics[width=\textwidth]{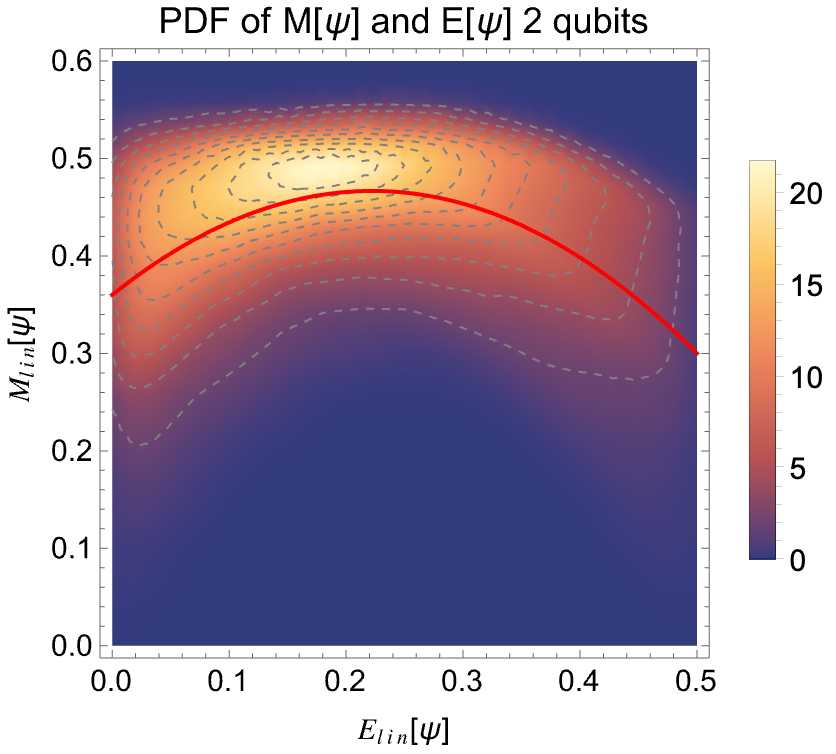}
        \caption{Scatter plot $\mlin$ versus $\elin$ of $N_{\rm sample}=2\times 10^5$ Haar random pure states $\ket{\psi}$ for 2 qubits with the (1|2) bipartition. The solid (red) line is the average value $\Tilde{M}(e)$ on the Schmidt orbit as a function of $e$ Eq.~\eqref{eq:Me_2dB}.}
        \label{fig:EM_fixed2}
    \end{subfigure}
    \hfill
    \begin{subfigure}[b]{0.48\textwidth}
        \centering
        \includegraphics[width=\textwidth]{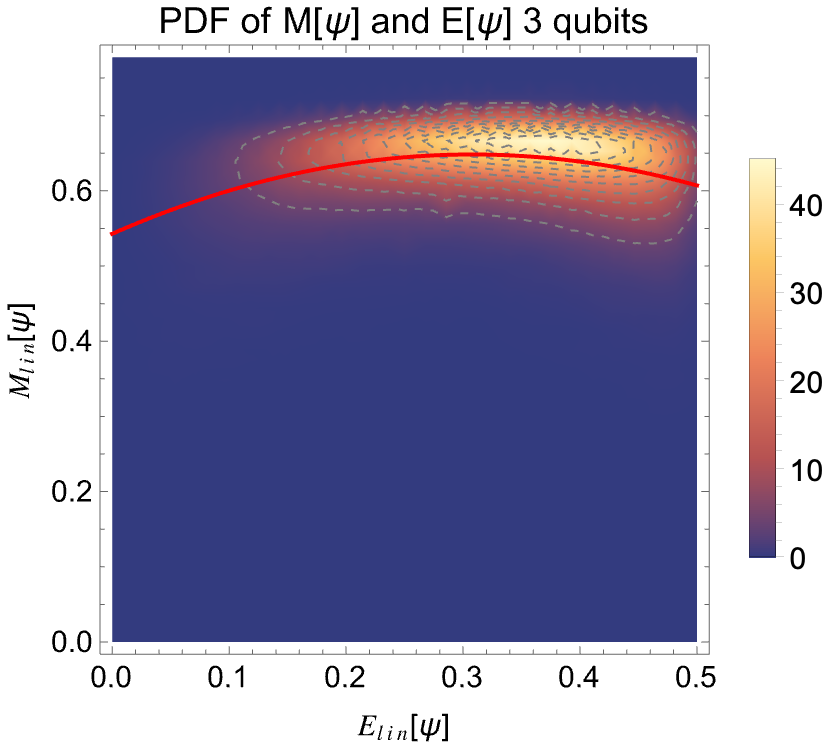}
        \caption{Scatter plot $\mlin$ versus $\elin$ of $N_{\rm sample}=3\times 10^5$ Haar random pure states $\ket{\psi}$ for 3 qubits with the (1|23) bipartition. The solid (red) line is the average value $\Tilde{M}(e)$ on the Schmidt orbit as a function of $e$ Eq.~\eqref{eq:Me_2dB}.}
        \label{fig:EM_fixed3}
    \end{subfigure}

    \vskip 0.5cm  

    \begin{subfigure}[b]{0.48\textwidth}
        \centering
        \includegraphics[width=\textwidth]{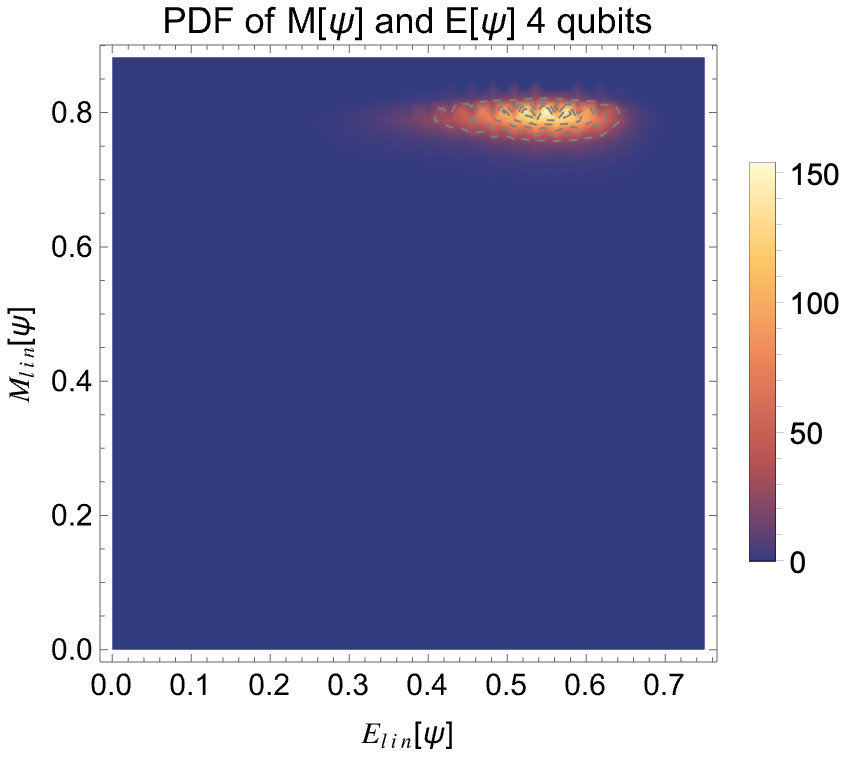}
        \caption{Scatter plot $\mlin$ versus $\elin$ of $N_{\rm sample}=4\times 10^5$ Haar random pure states $\ket{\psi}$ for 4 qubits with (12|34) half bipartition.}
        \label{fig:EM4}
    \end{subfigure}
    \hfill
    \begin{subfigure}[b]{0.48\textwidth}
        \centering
        \includegraphics[width=\textwidth]{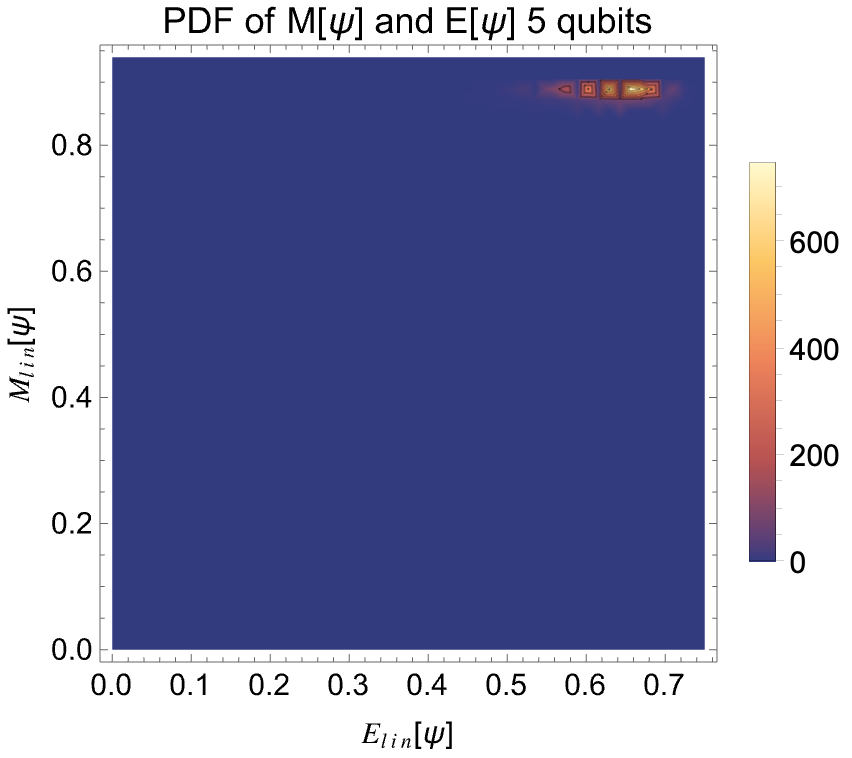}
        \caption{Scatter plot $\mlin$ versus $\elin$ of $N_{\rm sample}=10^6$ Haar random pure states $\ket{\psi}$ for 5 qubits with (12|345) bipartition.}
        \label{fig:EM5}
    \end{subfigure}
    \caption{Scatter plot $\mlin$ versus $\elin$ of Haar random pure states $\ket{\psi}$ for various system sizes, n=2,3,4,5.}
    \label{fig:EM_all}
\end{figure}

Using Eq.~\eqref{eq:m_e} one can readily obtain the leading behavior of the average magic at given entanglement in the above defined regions:
\begin{align}
\bar{M}(e) & =_{d\gg1}1-\frac{12e^{2}-24e+16}{d}+O\left(\frac{1}{d^{3/2}}\right)\,,\\
\bar{M}(e) & =_{d_{B}\gg d_{A}\gg1}1-\frac{12e^{2}-24e+16}{d}+O\left(\frac{1}{dd_{A}}\right)\,.
\end{align}
What the above results tell us is that the Schmidt orbits of magic are, at leading order, completely fixed by the entanglement. However, within these orbits, there is a finer structure where higher moments of the Schmidt spectrum are involved and carry information about the anti-flatness of the entanglement spectrum \cite{tirrito2024quantifying, cao2024gravitational}, see Fig.~\ref{fig:graphic} for an illustration.
 
In Fig.~\ref{fig:EM_all} we plot the full joint probability density up to 5 qubits for several bipartitions.

\subsection {Typicality of the result}
\label{sec:typicality}
The significance of the result obtained in Eq.~\eqref{eq:m_UAUB} is quantified by its typicality. 
To this end, the Chebyshev inequality is used, and in particular, the Bhatia-Davis \cite{bhatia2000BetterBoundVariance} inequality is employed to bound the variance of $\mlin(U_A\ot U_B\ket\psi)$. Such inequality states that, given a bounded random variable $x$,

\begin{equation}
    \Delta^2(X)\leq({\rm max}(X)-\exv(X))(\exv(X)-{\rm min}(X))
\end{equation}
Applying this inequality to $\mlin$, which is bounded between $0$ and $1$, this inequality reads

\begin{equation}
    \Delta^2[\mlin(U_A\ot U_B\ket\psi)]\leq \Bar{M}({\rm \lambda})^2-\Bar{M}({\rm \lambda})\,;
\end{equation}
direct evaluation of the right-hand side leads to the following bound:
\begin{equation}
    \begin{split}
         \Delta^2[\mlin(U_A\ot U_B\ket\psi)]&\leq  {\frac{16 d_A ^3 d_B ^3}{(d_A ^2-9)^2 (d_B ^2-9)^2} }= O\left(\frac{1}{d}\right)\,.
    \end{split}
\end{equation}
Hence, applying this bound to the Chebyshev inequality one can state that

\begin{equation}
    {\rm Pr}(|\mlin(U_A\ot U_B\ket\psi)-\Bar{M}({\bm \lambda})|\geq \epsilon)\leq O\left(\frac{1}{d \epsilon^2}\right)\,.
\end{equation}
Given that $0\leq \mlin(\ket\psi)\leq 1$, deviations can never scale larger than $O(1)$ and thus one can confidently say that the the average of $\mlin$ over $\slambda$ is typical with failure probability which exponentially vanishes as the number of qubits grows. Unfortunately,  the above bound does not allow us to place a useful bound on the relative fluctuations.

A stronger statement using Lévy typicality can be made.
One starts with the definition of a Lévy normal family as the family of metric measure spaces $(X^n,d^n)$ equipped with a probability Borel measure $\mu^n$ with $n\geq 1$ such that the concentration function reads
\begin{equation}
    \alpha_{(X^n, d^n,\mu^n)}(r) \leq C e^{-c n r^2}
\end{equation}
with constants $C,c>0$ and the following definitions
\begin{equation}
    \alpha_{(X^n, d^n,\mu^n)}(r)=\sup\left\{1- \mu(A_r); A \subset X, \mu(A)\geq \frac{1}{2}\right\}
\end{equation}
with $A_r:=\{ x \in X : d(x,A)< r\}$ the (open) $r$-neighborhood of $A$.
A peculiar case of a normal Lévy family is that of unit spheres.
Additionally, the product of normal Lévy families with the same $n$, equipped with the product measure and the direct sum of distances is again a normal Lévy family (Example 3.2 in \cite{milman1988heritage}).
For further details on the subject one can check \cite{ledoux2005ConcentrationMeasurePhenomenon}.
This property also extends in the case of different dimensions as long as the minimum of the two dimensions is large enough, i.e. $\min(d_A,d_B)\gg 1$.
It remains to show the Lipschitzianity of $\mlin(U_A\ot U_B\ket\psi)$ in this product space.
To do this, one can notice that the Lipschitzianity of $\Tr(Q\psi^{\ot 4})$ over the full Hilbert space, also implies Lipschitzianity on $\slambda$. To be specific, one can start from said Lipschitzianity, which reads:  
 \begin{equation}
   d|\Tr[Q(\psi^{\ot 4}-\phi^{\ot 4})]|\leq \frac{27}{5}\norm{\psi-\phi}_1\;,\forall\,\psi,\phi\;\text{pure states}\,.
 \end{equation}
 Now, picking two states from the same Schmidt orbit $\psil\,,\,\ket{\phi^{\bm \lambda}}\in\slambda$ and substituting them one gets
 \begin{equation}
     d|\Tr[Q(\psi^{{\bm \lambda}\ot 4}-\phi^{{\bm \lambda}\ot 4})]|\leq \frac{27}{5} \norm{\psi^{\bm \lambda}-\phi^{\bm \lambda}}_1\,.
 \end{equation} 
 Since the L\'evy lemma applies not only to the full Haar measure but to all functions defined on a spherical measure, it particularly applies to $\slambda$ for $\min(d_A,d_B)\gg 1$. Again, this does not allow us to argue about the smallness of relative fluctuations.
 
\section {Average magic at fixed entanglement of \texorpdfstring{$2\times d_B$}{TEXT}  bipartite random states}\label{example}
\begin{figure}[h!]
    \centering
    \includegraphics[width=0.50\linewidth]{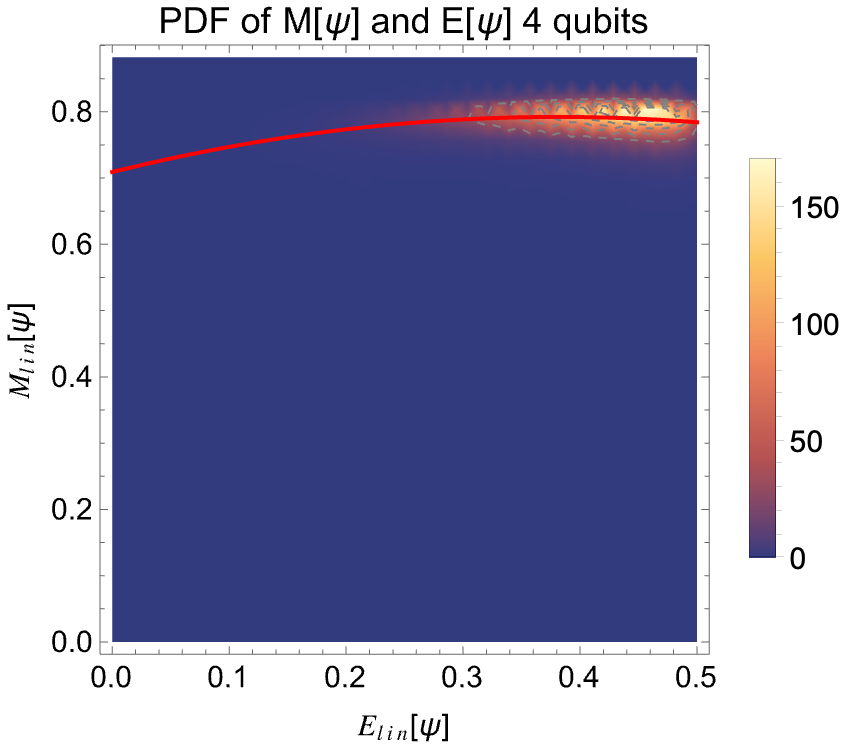}
    \caption{Scatter plot $\mlin$ versus $\elin$ of $N_{\rm sample}=4\times 10^5$ Haar random pure states $\ket{\psi}$ for 4 qubits with the (1|234) bipartition. The solid (red) line is the average value of $M_\mathrm{lin}$ at fixed entanglement $e$, $\Tilde{M}(e)$,  Eq.~\eqref{eq:Me_2dB}.}
    \label{fig:EM_fixed4}
\end{figure}

Following the approach in Ref.~\cite{giraud2007purity} we can compute the  \eqref{eq:m_e} for $d_A=2$ and a generic $d_B$: this reads 
\begin{equation}
\Tilde{M}_{2,d_B}(e)=\frac{5 d_B^2-24 d_B e^2+24 d_B e-d_B-60 e^2}{5 d_B (d_B+3)}\,, \label{eq:Me_2dB}
\end{equation}
and
\begin{equation}
    P_{2,d_B}(e)=\frac{d_B^2 e^{d_B-2}  \sqrt{1-2 e} \;  \Gamma \left(d_B+\frac{1}{2}\right) }{\sqrt{\pi } \; \Gamma (d_B-1)}\,. \label{eq:PE2_marginal}
\end{equation}
Numerical results for $d_B=2$, $d_B=4$, and $d_B=8$ are reported in Fig.~\ref{fig:EM_fixed2}, \ref{fig:EM_fixed3}, \ref{fig:EM_fixed4}. 

In the large $d_B$ limit the above equation becomes
\begin{equation}
    \Tilde{M}_{2,d_B}(e)=_{d \gg 1}1-\frac{8 \left(3 e^2-3 e+2\right)}{5 d_B}+O\left(\frac{1}{d_B^2}\right)\,,
\end{equation}
concluding that it approaches almost a constant value as $d_B$ grows.
One can recover the Haar average result by integrating over all possible values of entanglement
\begin{equation}
    \int_0^{\frac{1}{2}} de \; \Tilde{M}_{2,d_B}(e) \; P_{2,d_B}(e) = 1-\frac{4}{2 d_B+3}= \exv_U[\mlin (U \ket{\psi})]\,.
\end{equation}

\section{Entanglement-Magic duality through the lens of anti-flatness}
\label{ch:flat}
The result described in Proposition \ref{prop:M_AB} can be considered as follows from a generic perspective. A striking result in \cite{tirrito2024quantifying} is that a global quantity, $\mlin(\ket{\psi})$, is equal to a local spectral quantity for the reduced density operator. This spectral quantity is the average anti-flatness of the reduced density operator over the Clifford orbit. Define such anti-flatness as
\begin{equation}
    \mathcal{F}_A(\ket{\psi}):=\Tr(\psi_A^3)-\Tr^2(\psi_A^2)\,,
\end{equation}
with $\psi_A=\Tr_B(\ketbra{\psi})$. Then, the theorem in \cite{tirrito2024quantifying} states that
\begin{equation}
    \exv_{C}[\mathcal{F}_A(C \ket{\psi})]= \frac{(d^2-d_A^2)(d_A^2-1)}{(d^2-1)(d+2)d_A^2} \mlin(\ket{\psi})\,,
    \label{eq:mean_C}
\end{equation}
with $C \in \mathcal{C}(d)$ being Clifford unitaries and $\exv_{C}$ being the average over the Clifford group. Notice that the Clifford orbit preserves magic but changes the Schmidt coefficients.  {In words, the average anti-flatness over the magic preserving orbit is magic.}

We wonder if a dual statement is also true. Namely, we ask whether the average over the Schmidt orbit, that preserves the Schmidt coefficients but changes the magic yields a similar result, that is the anti-flatness of the states on the orbit.   {Namely, one would want to see if 
\begin{equation}
\bar{M}(\boldsymbol{\lambda}):=\mathbb{E}_{U_{A},U_{B}}\left[M_{\mathrm{lin}}(U_{A}\otimes U_{B}|\psi^{\bm{\lambda}}\rangle)\right] \overset{?}{=} \delta\mathcal F_A (\ket{\psi ^{\boldsymbol{\lambda}}}) +G(e(\ket{\psi^{\bm{\lambda}}}))\,,
\end{equation}
for a suitable constant $\delta$. In words,  the average magic over the entanglement preserving orbit is anti-flatness up to a universal function of entanglement $G(e)$.}

In order to more properly see such a duality between these statements we  rewrite Eq.~\eqref{eq:m_UAUB} as
\begin{equation}
\begin{split}
    \Bar{M}({\bm{\lambda}})
            &=\alpha+\beta e+\gamma e^2+\delta [f +(1-e)^2] +\mu \sum_{i} \lambda_i^4 
            \equiv \delta \mathcal{F}_A(\ket{\psi^{\bm{\lambda}}}) + G(e) + \sum_{i} \lambda_i^4\,,
   \end{split}
   \label{eq:M_e_f}
\end{equation}
where $e=1-\sum_i \lambda_i ^2$, $f=\sum_i \lambda_i ^3-\left( \sum_i \lambda_i ^2\right)^2 {= \mathcal{F}_A(\ket{\psi^{\bm{\lambda}}})}$ {, $G(e) := \alpha +\delta + ( \beta-2\delta) e +(\gamma+\delta) e^2$} and
\begin{equation}
    \begin{split}
        & \alpha = O(1), \quad \beta = O \left(\frac{1}{d}\right), \quad \gamma = O\left(\frac{1}{d}\right), \quad \delta= O\left(\frac{1}{d^2}\right) \quad \text{and} \quad \mu = O \left(\frac{d_A +d_B}{d^2} \right).
    \end{split}
\end{equation}
The above expression gives the mean value of magic upon having fixed the value of entanglement as a function of the anti-flatness and the entanglement itself apart from a term proportional to $\sum _{i} \lambda_i^4$.
As one notices, the average magic is not entirely determined by the state's  {entanglement and anti-flatness; the} supposed duality between entanglement and magic through anti-flatness, is thus broken  {by the term $\sum _{i} \lambda_i^4$}. A sign that such symmetry  {could not be} perfect can be seen in the fact the Clifford group allows one to reach states with maximum entanglement whereas the converse is not true: one cannot reach maximum magic states by means of factorized unitaries.

This collection of facts paints a picture in which the interplay between magic and entanglement is not symmetric, in the sense that magic needs entanglement to reach its maximal values, whereas entanglement does not need magic, due to the very relationship between the set of free operations of the two resources. The anti-flatness seems to capture some properties of both magic and entanglement, although imperfectly. 


\section {Conclusions and Outlook} 

This work establishes some technical results regarding the relationship between magic and entanglement in random states. We show that the two quantities are perfectly uncorrelated (in their linear versions) yet strongly dependent, which is a very non-trivial statistical result. 

We then compute the average magic on the Schmidt orbit, namely the average magic once the Schmidt coefficients of a bipartite state have been fixed. The result is that, in first approximation, the average over the Schmidt orbit is given by the linear entanglement, in second approximation, by the anti-flatness of the reduced density operator, plus other terms that break the duality between Clifford orbits and Schmidt orbits. Finally, we show that these results show typicality in the Hilbert space. 

In perspective, it would be important to deepen this analysis by finding exact results on the probability of magic conditioned to entanglement. These results will also constitute some useful tools for the study of the resource theory of non-local magic as well as the scrambling \cite{PhysRevLett.126.030601} and spreading of magic in local quantum systems. 
  

\section {Acknowledgments} 
The authors would like to thank G.~Ascione for clarifications about the concentration of measure for products of spheres.
Additionally, the authors recognize the fruitful discussions on the matter with S.~Cusumano and G.~Scuotto. This research was funded by the Research Fund for the Italian Electrical System under the Contract Agreement "Accordo di Programma 2022–2024" between ENEA and Ministry of the Environment and Energetic Safety (MASE)- Project 2.1 "Cybersecurity of energy systems". AH acknowledges support from the PNRR MUR project PE0000023-NQSTI and the PNRR MUR project CN 00000013-ICSC and stimulating conversation with G.~Zar\'and.



\appendix

\section{Haar integration over the flag manifold}
\label{sec:Haar}
If $H$ is a normal subgroup of a group $G$ one can define the quotient space $G/H$ which has also a group structure. If on $G$ there is a (unique) Haar measure $\mu_G(dg)$, this induces a Haar measure on $H$, $\mu_H(dh)$, and on the quotient $G/H$, $\mu_{G/H}(d\tilde{g})$. The measures satisfy the following relation

\begin{equation}
\int_{G/H}\mu_{G/H}(d\tilde{g})\int_{H}\mu_{H}(dh)f(\tilde{g}h)=\int_{G}\mu_{G}(dg)f(g)\, .
\end{equation}
In fact the above equation can be seen as a way to define a Haar measure on the quotient (see e.g.~\cite{nachbin_haar_1976}). 
We now use the above relation with $G=U(d_{A})\otimes U(d_{B})$ and
$H=\left[U(1)\right]^{d_{A}}$. The functions $f$ we are interested in, 
are $M_{\mathrm{lin}}\left(U_{A}\otimes U_{B}|\psi^{\boldsymbol{\lambda}}\rangle\right)$
and $E_{\mathrm{lin}}\left(U_{A}\otimes U_{B}|\psi^{\boldsymbol{\lambda}}\rangle\right)$.
Both of these functions are invariant under the action of $H$. In fact
$M_{\mathrm{lin}}$ is a function of the singular value decomposition (Schmidt decomposition) which is invariant under this symmetry while
the entanglement is only a function of the Schmidt coefficients $\boldsymbol{\lambda}$
(and not of the unitaries $U_{A}$ and $U_{B}$). 
Hence, for our case we have
\begin{equation}
\begin{split}
\int_{G/H}\mu_{G/H}(d\tilde{g})\int_{H}\mu_{H}(dh)f(\tilde{g}h) & =\int_{G/H}\mu_{G/H}(d\tilde{g})\int_{H}\mu_{H}(dh)f(\tilde{g})\\
 & =\int_{G/H}\mu_{G/H}(d\tilde{g})f(\tilde{g})=\int_{G}\mu_{G}(dg)f(g)\, ,
 \end{split}
\end{equation}
this means that we can replace Haar integration over the quotient
with Haar integration over the original group $U(d_{A})\otimes U(d_{B})$,
that is we obtain Eq.~\eqref{eq:prob_cond} of the main text. 


\section{Permutations conjugacy classes}
In this section a summary of the conjugacy classes of the symmetric groups of order 4, 6 and 8 is shown since they are heavily used in the computations of the variance of $\mlin$ and for the covariance between $\mlin$ and $\elin$.

\begin{table}[h]
    \centering
    \begin{tabular}{|c|c|}
        \hline
        \textbf{Cycle type} & \textbf{Size of conjugacy class} \\\hline   $()$ -- the identity element & 1\\ \hline   $(ab)$ & 6\\ \hline   $(ab)(cd)$& 3 \\ \hline   $(abc)$ & 8 \\ \hline  $(abcd)$ & 6\\
        \hline
    \end{tabular}
    \caption{Conjugacy classes of the symmetric group $S_4$ divided by cycle types.}
    \label{tab:S_4}
\end{table}

\begin{table}[h]
    \centering
    \begin{tabular}{|c|c|}
        \hline
        \textbf{Cycle type} & \textbf{Size of conjugacy class} \\
        \hline
         () -- the identity element & 1 \\
        \hline
         $(ab)$ & 15 \\
        \hline
         $(abc)$ & 40 \\
        \hline
        $(abcd)$ & 90 \\
        \hline
         $(ab)(cd)$ & 45 \\
        \hline
        $(abcde)$ & 144 \\
        \hline
        $(abc)(de)$ & 120 \\
        \hline
        $(ab)(cd)(ef)$ & 15 \\
        \hline
        $(abcd)(ef)$ & 90 \\
        \hline
        $(abc)(def)$ & 40 \\
        \hline
        $(abcdef)$ & 120 \\
        \hline
    \end{tabular}
    \caption{Conjugacy classes of the symmetric group $S_6$ divided by cycle types.}
    \label{tab:S_6}
\end{table}

\begin{table}[h]
    \centering
    \begin{tabular}{|c|c|}
        \hline
         \textbf{Representative element} & \textbf{Size of conjugacy class} \\
        \hline
        \((\,) \text{-- the identity element}\) & 1 \\
        \hline
         \((ab)\) & 28 \\
        \hline
         \((abc)\) & 112 \\
        \hline
         \((abcd)\) & 420 \\
        \hline
         \((ab)(cd)\) & 210 \\
        \hline
         \((abcde)\) & 1344 \\
        \hline
         \((abc)(de)\) & 1120 \\
        \hline
         \((abcdef)\) & 3360 \\
        \hline
         \((abcd)(ef)\) & 2520 \\
        \hline
         \((ab)(cd)(ef)\) & 420 \\
        \hline
         \((abc)(def)\) & 1120 \\
        \hline
         \((abcdefg)\) & 5760 \\
        \hline
         \((abc)(de)(fg)\) & 1680 \\
        \hline
         \((abcd)(efg)\) & 3360 \\
        \hline
         \((abcde)(fg)\) & 4032 \\
        \hline
         \((ab)(cd)(ef)(gh)\) & 105 \\
        \hline
        \((abcd)(ef)(gh)\) & 1260 \\
        \hline
         \((abc)(def)(gh)\) & 1120 \\
        \hline
         \((abcdef)(gh)\) & 3360 \\
        \hline
         \((abcde)(fgh)\) & 2688 \\
        \hline
         \((abcd)(efgh)\) & 1260 \\
        \hline
        \((abcdefgh)\) & 5040 \\
        \hline
    \end{tabular}
    \caption{Conjugacy classes of the symmetric group $S_8$ divided by cycle types.}
    \label{tab:S_8}
\end{table}


\section{On a peculiar pattern of traces and Pauli strings}
In the subsequent calculations, a lot of terms of the form $\Tr[(PP')^k]$ will be encountered: one can extract a rule to compute the sum over $P$ and $P'$.
\newline
Given two Pauli strings $P$ and $P'$, then 
    \begin{equation}
         \sum_{P, P'} \Tr[(PP')^k] = \begin{cases} d^5 \quad k=4p\\
         d^3 \quad \text{otherwise}
         \end{cases}=\Big(14 + 6 (-1)^k + 12 \cos \Big(\frac{\pi}{2}k\Big) \Big)^n\,,
    \end{equation}
for $d=2^n$, $n \in \mathbb{N}$ the number of qubits and $k,p \in \mathbb{N}_+$.
\newline
One can start by writing
\begin{equation}
    \sum_{P, P' \in \pauli_n} \Tr[(PP')^k]= \sum_{\alpha ,\beta} \prod_{i=1}^n \Tr[(\sigma^{\alpha_i} \sigma^{\beta_i})^k]\, ,
\end{equation}
where  $P=\sigma^{\alpha_1} \otimes \ldots \otimes \sigma^{\alpha_n}$ and $P'=\sigma^{\beta_1} \otimes \ldots \otimes \sigma^{\beta_n}$ with $\alpha,\beta \in\{0,1,2,3\}^{\times n}$ that defines the Pauli matrices.
Now define 
\begin{equation}
    M_{\alpha,\beta}^{(k)}:= \Tr[(\sigma^{\alpha} \sigma^{\beta})^k]\, ,
\end{equation}
then
\begin{equation}
    \sum_{P,P' \in \pauli_n} \Tr[(PP')^k] =\Big(\sum_{\alpha \beta} M_{\alpha,\beta}^{(k)}\Big)^n \, .
\end{equation}
A close formula form can be written for $M_{\alpha,\beta}$, namely all the combinations of Pauli matrices reads
\begin{equation}
    \sigma^\alpha \sigma^\beta =
    \begin{bmatrix}
        \id & X & Y & Z \\
        X & \id & i Z & -i Y\\
        Y & -i Z & \id & i X \\
        Z &i Y & -iX& \id
    \end{bmatrix}\,,
\end{equation}
and the trace of the $k$-power
\begin{equation}
    \tr\left[\left(\sigma^{\alpha}\sigma^{\beta}\right)^{k}\right]=2 
    \begin{bmatrix}
        1 & \delta_{k,{\rm even}} & \delta_{k,{\rm even}} & \delta_{k,{\rm even}}\\
  \delta_{k,{\rm even}} & 1 & \left(i\right)^{k}\delta_{k,{\rm even}} & \left(-i\right)^{k}\delta_{k,{\rm even}}\\
\delta_{k,{\rm even}} & \left(- i\right)^{k}\delta_{k,{\rm even}} & 1 & \left(i\right)^{k}\delta_{k,{\rm even}}\\
\delta_{k,{\rm even}} & \left(i\right)^{k}\delta_{k,{\rm even}}  & \left(- i\right)^{k}\delta_{k,{\rm even}} & 1
    \end{bmatrix}\,,
\end{equation}
where $\delta_{k,{\rm even}}=\frac{1+(-1)^k}{2}$ and the fact that $\Tr[(\sigma^\gamma)^k]= 2 \delta_{k,{\rm even}}$ for $\gamma\in \{1,2,3\}$.
Finally one has
\begin{equation}
    \sum_{\alpha \beta} M_{\alpha,\beta}^{(k)}= 2\left[4+ 6 \delta_{k,{\rm even}} + 6 \cos \Big(\frac{\pi}{2}k\Big) \right] =14 + 6 (-1)^k + 12 \cos \Big(\frac{\pi}{2}k\Big) \,.
\end{equation}
A simple check can be made for $k=2$, namely
\begin{align}
    \sum_{P,P' \in \pauli_n} \Tr[(PP')^2] &= \sum_{P,P' \in \pauli_n} \Tr[PP'PP']\\
    &=  \Tr[\perm_2 \Big(\sum_P P^{\otimes 2} \Big) \Big(\sum_{P'} P'^{\otimes 2}\Big)]\\
    &= d^2 \Tr[\perm_2 \perm_2  \perm_2 ]\\
    &= d^2 \Tr[\perm_2]= d^3\, ,
\end{align}
with $\perm_2$ the swap over the two copies.
Extending the strategy further by computing 
\begin{equation}
         \sum_{P_1, P_2, P_3} \Tr[(P_1 P_2 P_3)^k] = \begin{cases} d^2 \,5^{\log_2 d} \quad k=2p+1\\
         d^7 \quad \quad \quad \,\, k=4p\\
         d^5 \quad \quad \quad \,\, \text{otherwise}
         \end{cases}=\Big( 50 + 30 (-1)^k + 48 \cos \Big( \frac{\pi}{2}k\Big) \Big)^n \,,
\end{equation} 
\begin{equation}
         \sum_{P_1, P_2, P_3, P_4} \Tr[(P_1 P_2 P_3 P_4)^k] = \begin{cases} d^3 \, 7^{\log_2 d} \quad k=2p+1\\
         d^9 \quad \quad \quad \, \, \, k=4p\\
         d^5 \quad \quad \quad \,\, \, \text{otherwise}
         \end{cases}= \Big(164 + 108 (-1)^k + 240\cos \Big( \frac{\pi}{2}k\Big) \Big)^n\,, 
\end{equation}
for $d=2^n$, $n \in \mathbb{N}$ the number of qubits and $k,p \in \mathbb{N}_+$.


\section{Explicit calculations for the SE variance}\label{var_calc}
Now we show the explicit calculations for the fluctuations of $\mlin$  , i.e.
\begin{equation}
    \mlin(\psi)= 1- d \Tr[Q \psi^{\otimes 4}]\, ,
\end{equation}
with $Q=\frac{1}{d^2}\sum_{P\in \pauli_n} P^{\otimes 4}$.
The fluctuations $\Delta^2(\mlin)$ will be computed according to the Haar measure
\begin{equation}
    \exv_U[(\mlin(\psi_U)- \exv_U[\mlin(\psi_U)])^2]= d^2\, \exv[\Tr^2[\psi_U ^{\otimes 4} Q]]- \Big(\frac{4}{d+3} \Big)^2\, ,
\end{equation}
using $\exv_U[\mlin(\psi_U)]= 1 - \frac{4}{d+3}$ \cite{leone2022stabilizer}.
The main calculation revolves around
\begin{equation}
    \exv_U[\Tr^2[\psi_U ^{\otimes 4} Q]]= \Tr[\exv_U[\psi_U ^{\otimes 8}](Q \otimes Q)] \, .
    \label{eq:Q_Qterm}
\end{equation}
The authors in \cite{roberts2017ChaosComplexityDesign,mele2024introduction} show that the Haar average of $k$ copies of a pure density operator is given by
\begin{equation}\label{st_avg}
    \exv_U[\psi_U ^{\otimes k}]= \frac{1}{\prod_{i=0}^{k-1} (d+i)}\sum_{\pi \in S_k} \perm_{\pi} \, ,
\end{equation}
where the sum runs over the elements of the symmetric group of order $k$.
Substituting this expression for $k=8$ in Eq.~\eqref{eq:Q_Qterm} and explicit expansion of $Q$ yields the following:
\begin{equation}\label{Q_Q_expanded}
    \exv_U[\Tr^2[\psi_U ^{\otimes 4} Q]]=\frac{1}{d^2\prod_{i=0}^7 (d+i)}\sum_{\pi\in S_8}\sum_{P,P'}\Tr[(P^{\ot 4} \ot P'^{\ot 4})\perm_\pi]\,.
\end{equation}
In order to compute the traces in \eqref{Q_Q_expanded}, the following property of permutation operators has been used \cite{oliviero2021RandomMatrixTheory}:
\begin{equation}
    \Tr\left[\perm_\pi \bigotimes_{j=1}^k A_j\right]=\prod_{j=1}^r \Tr\left(\prod_{l=1}^{k_j} A_{\tau_j(l)}\right)\,,
\end{equation}
where $\perm_\pi$ is a permutation comprised of $r$ cycles $\tau_j$, each of length $k_j$ respectively. However, the numbers of elements of $S_8$ in Eq.~\eqref{Q_Q_expanded} renders a one-by-one manual evaluation unfeasible: the calculation will be approached by dividing the permutations in conjugacy classes/cycle types, using the data from Table \ref{tab:S_8}. The passages are shown below:

$\bullet \; 1-$ cycles
\vspace{-1pt}
\begin{equation}
    \begin{split}
        d^8\sum_{P,P'\in \pauli}\delta_{P,\id}\delta_{P',\id}=d^8\,.
    \end{split}
\end{equation}

$\bullet \; 2-$ cycles
\vspace{-1pt}
\begin{equation}
    \begin{split}
        \sum_{P,P'\in \pauli}&6 d \Tr[P]^4 \Tr[P']^2+6 d \Tr[P]^2 \Tr[P']^4+16 \Tr[P]^3 \Tr[P']^3 \Tr[P'P]=28d^7\,.
    \end{split}
\end{equation}

$\bullet \; 3-$ cycles
\vspace{-1pt}
\begin{equation}
    \begin{split}
        \sum_{P,P'\in \pauli}56 \Tr[P]^4 \Tr[P']^2+56 \Tr[P]^2 \Tr[P']^4=112 d^6\,.
    \end{split}
\end{equation}

$\bullet \, 4-$ cycles
\vspace{-1pt}
\begin{equation}
    \begin{split}
      \sum_{P,P' \in \pauli}& 144 d \Tr[P]^2 \Tr[P']^2+6 d \Tr[P]^4+6 d \Tr[P']^4+96 \Tr[P]^3 \Tr[P'] \Tr[PP']\\&+72 \Tr[P]^2 \Tr[P']^2 \Tr[PP'PP']+96 \Tr[P] \Tr[P']^3 \Tr[PP']+324 d^2=408 d^5 + 12 d^7\,.
    \end{split}
\end{equation}

$\bullet \, 5-$ cycles
\vspace{-1pt}
\begin{equation}
    \begin{split}
        \sum_{P,P'\in \pauli}&1152 \Tr[P]^2 \Tr[P']^2+96 \Tr[P]^4+96 \Tr[P']^4=1152 d^4 + 192 d^6\,.
    \end{split}
\end{equation}

$\bullet \, 6-$ cycles
\vspace{-1pt}
\begin{equation}
    \begin{split}
     \sum_{P,P'\in \pauli}&432 d^2 \Tr[P]+432 d^2 \Tr[P']+288 d \Tr[P] \Tr[PP'PP']+288 d \Tr[P'] \Tr[PP'PP']\\&+1728 \Tr[P] \Tr[P'] \Tr[PP']+192 \Tr[P] \Tr[P'] \Tr[PP'PP'PP']=1920 d^3 + 1440 d^5\,.
    \end{split}
\end{equation}

$\bullet \, 7-$ cycles
\vspace{-1pt}
\begin{equation}
    \begin{split}
        \sum_{P,P'\in \pauli}&2880 \Tr[P]^2+2880 \Tr[P']^2=5760 d^4\,.
    \end{split}
\end{equation}

$\bullet \, 8-$ cycles
\vspace{-1pt}
\begin{equation}
    \begin{split}
        \sum_{P,P'\in \pauli}&2304 \Tr[PP'PP']+144 \Tr[PP'PP'PP'PP']+2592 d=2304 d^3 + 2736 d^5\,.
    \end{split}
\end{equation}

$\bullet \, (ab)(cd)-$ cycles
\vspace{-1pt}
\begin{equation}
    \begin{split}
        \sum_{P,P'\in \pauli}&36 d^2 \Tr[P]^2 \Tr[P']^2+3 d^2 \Tr[P]^4+3 d^2 \Tr[P']^4+48 d \Tr[P]^3 \Tr[P'] \Tr[PP']\\&+48 d \Tr[P] \Tr[P']^3 \Tr[PP']+72 \Tr[P]^2 \Tr[P']^2 \Tr[PP']^2=204 d^6 + 6 d^8\,.
    \end{split}
\end{equation}

$\bullet \, (abc)(de)-$ cycles
\vspace{-1pt}
\begin{equation}
    \begin{split}
        \sum_{P,P'\in \pauli}&384 d \Tr[P]^2 \Tr[P']^2+48 d \Tr[P]^4+48 d \Tr[P']^4+320 \Tr[P]^3 \Tr[P'] \Tr[PP']\\&+320 \Tr[P] \Tr[P']^3 \Tr[PP']=1024 d^5 + 96 d^7\,.
    \end{split}
\end{equation}

$\bullet \, (abcd)(ef)-$ cycles
\vspace{-1pt}
\begin{equation}
    \begin{split}
        \sum_{P,P'\in \pauli}&180 d^2 \Tr[P]^2+180 d^2 \Tr[P']^2+72 d \Tr[P]^2 \Tr[PP'PP']+1152 d \Tr[P] \Tr[P'] \Tr[PP']\\&+72 d \Tr[P']^2 \Tr[PP'PP']+288 \Tr[P]^2 \Tr[PP']^2+288 \Tr[P] \Tr[P'] \Tr[PP'] \Tr[PP'PP']\\&+288 \Tr[P']^2 \Tr[PP']^2=2016 d^4 + 504 d^6\,.
    \end{split}
\end{equation}

$\bullet \, (ab)(cd)(ef)-$ cycles
\vspace{-1pt}
\begin{equation}
    \begin{split}
        \sum_{P,P'\in \pauli}&18 d^3 \Tr[P]^2+18 d^3 \Tr[P']^2+144 d^2 \Tr[P] \Tr[P'] \Tr[PP']+72 d \Tr[P]^2 \Tr[PP']^2\\&+72 d \Tr[P']^2 \Tr[PP']^2+96 \Tr[P] \Tr[P'] \Tr[PP']^3=384 d^5 + 36 d^7\,.
    \end{split}
\end{equation}

$\bullet \, (abc)(def)-$ cycles
\vspace{-1pt}
\begin{equation}
    \begin{split}
        \sum_{P,P'\in \pauli}&832 \Tr[P]^2 \Tr[P']^2+144 \Tr[P]^4+144 \Tr[P']^4=832 d^4 + 288 d^6\,.
    \end{split}
\end{equation}

$\bullet \, (abc)(de)(fg)-$ cycles
\vspace{-1pt}
\begin{equation}
    \begin{split}
        \sum_{P,P'\in \pauli}&168 d^2 \Tr[P]^2+168 d^2 \Tr[P']^2+768 d \Tr[P] \Tr[P'] \Tr[PP']+288 \Tr[P]^2 \Tr[PP']^2\\&+288 \Tr[P']^2 \Tr[PP']^2=336 d^6+1344 d^4\,.
    \end{split}
\end{equation}

$\bullet \, (abcd)(efg)$
\vspace{-1pt}
\begin{equation}
    \begin{split}
        \sum_{P,P'\in \pauli} &624 d^2 \Tr[P] + 624 d^2 \Tr[P'] + 1536 \Tr[P] \Tr[P'] \Tr[P . P'] + 
 288 d \Tr[P] \Tr[P  P'  P  P']\\ &+ 288 d \Tr[P'] \Tr[P  P'  P  P']=1824 d^5+1536 d^3\,.
    \end{split}
\end{equation}

$\bullet \, (abcde)(fg)-$ cycles
\vspace{-1pt}
\begin{equation}
    \begin{split}
        \sum_{P,P'\in \pauli}&864 d \Tr[P]^2+864 d \Tr[P']^2+2304 \Tr[P] \Tr[P'] \Tr[PP']=2304 d^3 + 1728 d^5\,.
    \end{split}
\end{equation}

$\bullet \, (ab)(cd)(ef)(gh)-$ cycles
\vspace{-1pt}
\begin{equation}
    \begin{split}
        \sum_{P,P'\in \pauli}&72 d^2 \Tr[PP']^2+24 \Tr[PP']^4+9 d^4=96 d^6 + 9 d^8\,.
    \end{split}
\end{equation}

$\bullet \, (abcd)(ef)(gh)-$ cycles
\vspace{-1pt}
\begin{equation}
    \begin{split}
        \sum_{P,P'\in \pauli}&72 d^2 \Tr[PP'PP']+864 d \Tr[PP']^2+144 \Tr[PP']^2 \Tr[PP'PP']+180 d^3=180 d^7+1080 d^5\,.
    \end{split}
\end{equation}

\vspace{-1pt}
\begin{equation}
    \begin{split}
        \sum_{P,P'\in \pauli}&240 d \Tr[P]^2+240 d \Tr[P']^2+640 \Tr[P] \Tr[P'] \Tr[PP']=640 d^3 + 480 d^5\,.
    \end{split}
\end{equation}

$\bullet \, (abcdef)(gh)-$ cycles
\vspace{-1pt}
\begin{equation}
    \begin{split}
        \sum_{P,P'\in \pauli}&576 d \Tr[PP'PP']+1728 \Tr[PP']^2+192 \Tr[PP'] \Tr[PP'PP'PP']+864 d^2=2496 d^4 + 864 d^6\,.
    \end{split}
\end{equation}

$\bullet \, (abcde)(fgh)-$ cycles
\vspace{-1pt}
\begin{equation}
    \begin{split}
        \sum_{P,P'\in \pauli}&1344 \Tr[P]^2+1344 \Tr[P']^2=2688 d^4\,.
    \end{split}
\end{equation}

$\bullet \, (abcd)(efgh)-$ cycles
\vspace{-1pt}
\begin{equation}
    \begin{split}
        \sum_{P,P'\in \pauli}&288 d \Tr[PP'PP']+576 \Tr[PP']^2+72 \Tr[PP'PP']^2+324 d^2=864 d^4 + 396 d^6\,.
    \end{split}
\end{equation}

Summing every piece, one get

\begin{equation}
    \begin{split}
        \exv_\psi(\Tr[(Q\ot Q)\psi^{\ot 8}])=\frac{16 \left(d^2+15 d+68\right)}{(d+3) (d+5) (d+6) (d+7)}\,.
    \end{split}
\end{equation}

Hence the final result of the variance of $\mlin(\psi)$ reads

\begin{equation}
    \Delta^2 \mlin(\psi)=\frac{96 (d-1)}{(d+3)^2 (d+5) (d+6) (d+7)}= O\left(\frac{1}{d^4}\right)\,.
\end{equation}

\section{Explicit calculation of the covariance between \texorpdfstring{$\mlin$}{TEXT} and \texorpdfstring{$\elin$}{TEXT}}\label{cov_calc}
The object of this section is to show explicit calculations for ${\rm cov}(\mlin,\elin)$. The object reads
\begin{equation}
    \begin{split}
        {\rm Cov}(\mlin,\elin)&=\exv_U[\mlin(\psi_U)\elin(\psi_U)]-\exv_U[\mlin(\psi_U)]\exv_U[\elin(\psi_U)]\\
        &=d\exv_U[\Tr[(\psi_U ^A) ^2]\Tr[Q\psi^{\ot 4}]]-\frac{4(d_A+d_B)}{(d_A d_B+1)(d_A d_B+3)}\,.
    \end{split}
\end{equation}  

The main calculation revolves around
\begin{equation}
\label{eq:cor_interesting}
   \exv_U[\Tr[(\psi_U ^A)^2]\Tr[Q \psi_U ^{\otimes 4}]] 
   = \Tr[\exv_U[\psi^{\otimes 6}](\perm_2 ^A \otimes \id_B ^{\ot 2}) \otimes Q] 
   \, .
\end{equation}
Substituting the expression from Eq.~\eqref{st_avg} for $k=6$ the formula reads
\begin{equation}
    \begin{split}
        \exv_U&[\Tr[(\psi_U ^A)^2]\Tr[Q \psi_U ^{\otimes 4}]]=\frac{1}{\prod_{i=0}^5 (d+i)}\sum_{\pi\in S_6} \Tr\{\perm_\pi[(\perm_2 ^A \otimes \id_B ^{\ot 2}) \otimes Q]\}\\
        &=\frac{1}{d^2 d_A\prod_{i=0}^5 (d+i)}\sum_{\pi\in S_6}\sum_{P_A \in \pauli_{n_A}}\sum_{P_A' \in \pauli_{n_A},P_B' \in \pauli_{n_B}}\Tr\{\perm_\pi[(P_A ^{\ot 2}\ot \id_B ^{\ot 2})\ot(P_A' \ot P_B')^{\ot 4}]\}\,,
    \end{split}
\end{equation}
where $\perm_2 ^A=\frac{1}{d_A}\sum_{P_A \in \pauli_A}P_A ^{\ot 2}$.
Moreover, using the fact that $\perm_\pi=\perm_\pi ^A \ot \perm_\pi ^B$, one can further write
\begin{equation}
    \begin{split}
        \exv_U[\Tr[(\psi_U ^A)^2]\Tr[Q \psi_U ^{\otimes 4}]]&=\frac{1}{d^2 d_A\prod_{i=0}^5 (d+i)}\sum_{\pi\in S_6}\sum_{P_A,P_A' \in \pauli_{n_A}}\Tr[\perm_\pi ^A (P_A ^{\ot 2}\ot P_A'^{\ot 4})]\\&\sum_{P_B' \in \pauli_{n_B}}\Tr[\perm_\pi ^B (P_B' {}^{\ot 4}\ot \id_B ^{\ot 2})]\,.
    \end{split}
\end{equation}
Pulling the data from Table \ref{tab:S_6}, one computes the sums over permutations according to conjugacy class: calculations follow below.

$\bullet \,$- Identity
\vspace{-1pt}
\begin{equation}
	\begin{split}
		d_A ^6 d_B ^6\sum_{P_A,P_A ',P_B}\delta_{P_A,\id_A}\delta_{P_A',\id_A}\delta_{P_A',\id_A}=d_A ^6 d_B ^6\,.
	\end{split}
\end{equation}

$\bullet \, 2$- cycles
\vspace{-1pt}
\begin{equation}
	\begin{split}
		\sum_{P_A,P_A ',P_B}&6 d_A  d_B ^3 \Tr[P_A]^2 \Tr[P_{A}' ]^2 \Tr[P_B]^2+d_A  d_B \Tr[P_{A}' ]^4 \Tr[P_B]^4\\&+8 d_B \Tr[P_A] \Tr[P_{A}' ]^3 \Tr[P_B]^4 \Tr[P_{A}' P_A]=d_A ^7 d_B ^5+14 d_A ^5 d_B ^5\,.
	\end{split}
\end{equation}

$\bullet \,3$- cycles
\vspace{-1pt}
\begin{equation}
	\begin{split}
		\sum_{P_A,P_A ',P_B}&32 d_B^2 \Tr[P_A]^2 \Tr[P_{A}' ]^2 \Tr[P_B]^2+8 \Tr[P_{A}' ]^4 \Tr[P_B]^4=8 d_A ^6 d_B ^4+32 d_A ^4 d_B ^4\,.
	\end{split}
\end{equation}

$\bullet \,4$- cycles
\vspace{-1pt}
\begin{equation}
	\begin{split}
		\sum_{P_A,P_A ',P_B}&6 d_A  d_B ^3 \Tr[P_A]^2+24 d_A  d_B \Tr[P_{A}' ]^2 \Tr[P_B]^2+16 d_B \Tr[P_A] \Tr[P_{A}' ] \Tr[P_B]^2 \Tr[P_A P_{A}' ]\\&+32 d_B \Tr[P_A] \Tr[P_{A}' ] \Tr[P_B]^2 \Tr[P_{A}' P_A]+6 d_B \Tr[P_{A}' ]^2 \Tr[P_B]^2 \Tr[P_A P_{A}' P_A P_{A}' ]\\&+6 d_B \Tr[P_{A}' ]^2 \Tr[P_B]^2 \Tr[P_{A}' P_A P_{A}' P_A]=6 d_A ^5 d_B ^5+36 d_A ^5 d_B ^3+48 d_A ^3 d_B ^3\,.
	\end{split}
\end{equation}

$\bullet \,5$- cycles
\vspace{-1pt}
\begin{equation}
	\begin{split}
		\sum_{P_A,P_A ',P_B}&48 d_B^2 \Tr[P_A]^2+8 \Tr[P_{A}' ] \Tr[P_B]^2 \Tr[P_{A}' P_A P_{A}' P_A P_{A}' ]+88 \Tr[P_{A}' ]^2 \Tr[P_B]^2\\&=48 d_A ^4 d_B ^4+96 d_A ^4 d_B^2\,.
	\end{split}
\end{equation}

$\bullet \,6$- cycles
\vspace{-1pt}
\begin{equation}
	\begin{split}
		\sum_{P_A,P_A ',P_B}&24 d_B \Tr[P_A P_{A}' P_A P_{A}' ]+24 d_B \Tr[P_{A}' P_A P_{A}' P_A]+72 d_A  d_B=72 d_A ^5 d_B ^3+48 d_A ^3 d_B ^3\,.
	\end{split}
\end{equation}

$\bullet \,(ab) (dc)$- cycles
\vspace{-1pt}
\begin{equation}
	\begin{split}
		\sum_{P_A,P_A ',P_B}&3 d_A ^2 d_B ^4 \Tr[P_A]^2+6 d_A ^2 d_B^2 \Tr[P_{A}' ]^2 \Tr[P_B]^2\\&+24 d_A  d_B^2 \Tr[P_A] \Tr[P_{A}' ] \Tr[P_B]^2 \Tr[P_{A}' P_A]+12 \Tr[P_{A}' ]^2 \Tr[P_B]^4 \Tr[P_{A}' P_A]^2\\&=3 d_A ^6 d_B ^6+6 d_A ^6 d_B ^4+36 d_A ^4 d_B ^4\,.
	\end{split}
\end{equation}

$\bullet \,(abc) (de) $- cycles
\vspace{-1pt}
\begin{equation}
	\begin{split}
		\sum_{P_A,P_A ',P_B}&24 d_A  d_B ^3 \Tr[P_A]^2+32 d_A  d_B \Tr[P_{A}' ]^2 \Tr[P_B]^2\\&+64 d_B \Tr[P_A] \Tr[P_{A}' ] \Tr[P_B]^2 \Tr[P_{A}' P_A]=24 d_A ^5 d_B ^5+32 d_A ^5 d_B ^3+64 d_A ^3 d_B ^3\,.
	\end{split}
\end{equation}

$\bullet \,(abcd) (ef)$- cycles
\vspace{-1pt}
\begin{equation}
	\begin{split}
		\sum_{P_A,P_A ',P_B}&6 d_A  d_B^2 \Tr[P_A P_{A}' P_A P_{A}' ]+6 d_A  d_B^2 \Tr[P_{A}' P_A P_{A}' P_A]+32 \Tr[P_B]^2 \Tr[P_{A}' P_A]^2\\&+16 \Tr[P_B]^2 \Tr[P_A P_{A}' ] \Tr[P_{A}' P_A]+30 d_A ^2 d_B^2=30 d_A ^6 d_B ^4+12 d_A ^4 d_B ^4+48 d_A ^4 d_B^2\,.
	\end{split}
\end{equation}

$\bullet \, (ab)(cd)(ef)$- cycles
\vspace{-1pt}
\begin{equation}
	\begin{split}
		\sum_{P_A,P_A ',P_B}&12 d_A  d_B \Tr[P_B]^2 \Tr[P_{A}' P_A]^2+3 d_A ^3 d_B ^3=3 d_A ^7 d_B ^5+12 d_A ^5 d_B ^3\,.
	\end{split}
\end{equation}

$\bullet \,(abc)(def)$- cycles
\vspace{-1pt}
\begin{equation}
	\begin{split}
		\sum_{P_A,P_A ',P_B}&24 d_B^2 \Tr[P_A]^2+16 \Tr[P_{A}' ]^2 \Tr[P_B]^2=24 d_A ^4 d_B ^4+16 d_A ^4 d_B^2\,.
	\end{split}
\end{equation}

Summing all those pieces together, one gets 
\begin{equation}
     \exv_U[\Tr[(\psi_U ^A)^2]\Tr[Q \psi_U ^{\otimes 4}]] 
   =\frac{4 (d_A + d_B)}{d_A ^4 d_B ^3 ( d_A d_B+1 ) (d_A d_B+3)}\,,
\end{equation}
and finally
\begin{equation}
    {\rm Cov}(\mlin,\elin)=\frac{4 (d_A+d_B)}{d_A d_B (d_A d_B+1) (d_A d_B+3)}-\frac{4 (d_A+d_B)}{d_A d_B (d_A d_B+1) (d_A d_B+3)}=0\,.
\end{equation}


 {\section{Random variables and correlations}
\label{App:example_correlation}
Consider two independent random variables normally distributed as
\begin{equation}
    Z_1\,,\,Z_2 \sim \mathcal{N}\left(\mu=0,\sigma_n=\frac{1}{n}\right)\,,
\end{equation}
so that
\begin{equation}
    \mathcal{N}\left(\mu=0,\sigma_n = \frac{1}{n}\right) = \frac{e^{-\frac{x^2}{2 \sigma_n^2}}}{\sqrt{2 \pi \sigma_n}} \stackrel{n \to \infty }{\longrightarrow} \delta(x)\,.
\end{equation}
Then, if we define
\begin{equation}
    X= Z_1\sim \mathcal{N}\left(\mu=0,\sigma_n=\frac{1}{n}\right)\,, \quad Y= \rho Z_1 + \sqrt{1-\rho^2} Z_2 \sim \mathcal{N}\left(\mu=0,\sigma_n=\frac{1}{n}\right)\,,
\end{equation}
with $\rho\in(-1,1)$, the covariance between $X$ and $Y$ reads
\begin{equation}
    \begin{split}
        {\rm Cov}(X,Y)&= \exv[XY]-\exv[X] \exv[Y]\\
        &= \exv[XY] = \rho \,\exv[Z_1^2]+ \sqrt{1- \rho^2} \,\exv[Z_1 Z_2]\\
        &= \rho \,\exv[Z_1^2]=\frac{\rho}{n^2} \to_{n \to \infty }0\,.
    \end{split}
\end{equation}
However, if we want to see whether these two quantities are uncorrelated in the large $n$ limit we need to compute
\begin{equation}
    {\rm Corr}(X,Y)=\frac{{\rm Cov}(X,Y)}{\sqrt{\Delta^2 X \, \Delta^2 Y }}= \frac{\rho/n^2}{\sqrt{1/n^4}}=\rho\,.
\end{equation}
which does not scale with $n$.}


 {
\section{Gaussian approximation}
\label{app:gauss_approx}
In this section, we report the numerical evidence for the goodness of the approximation of linear magic and entanglement joint probability distribution.
We extracted $N_{sample}=10^5$ Haar random states $\ket{\psi}$ and then compute the $\mlin(\ket{\psi})$ and $\elin(\ket{\psi})$ for half bipartition, i.e. the number of qubits of the subsystem A is always given by $n_A=\left\lfloor \frac{n}{2} \right\rfloor$, with $n$ the total number of qubits.
For each system we compute the Kullback–Leibler divergence and the $L_1$ norm between the numerical PDF and the Gaussian approximation in \ref{eq:gauss_approx}, see Fig.\ref{fig:KL_L1_decay}. 
We see an exponential decay, showed as exponential fit in \ref{fig:KL_L1_decay}, of the distances defined above in the large $d=2^n$ limit.
\begin{figure}
    \centering
    \includegraphics[width=0.5\linewidth]{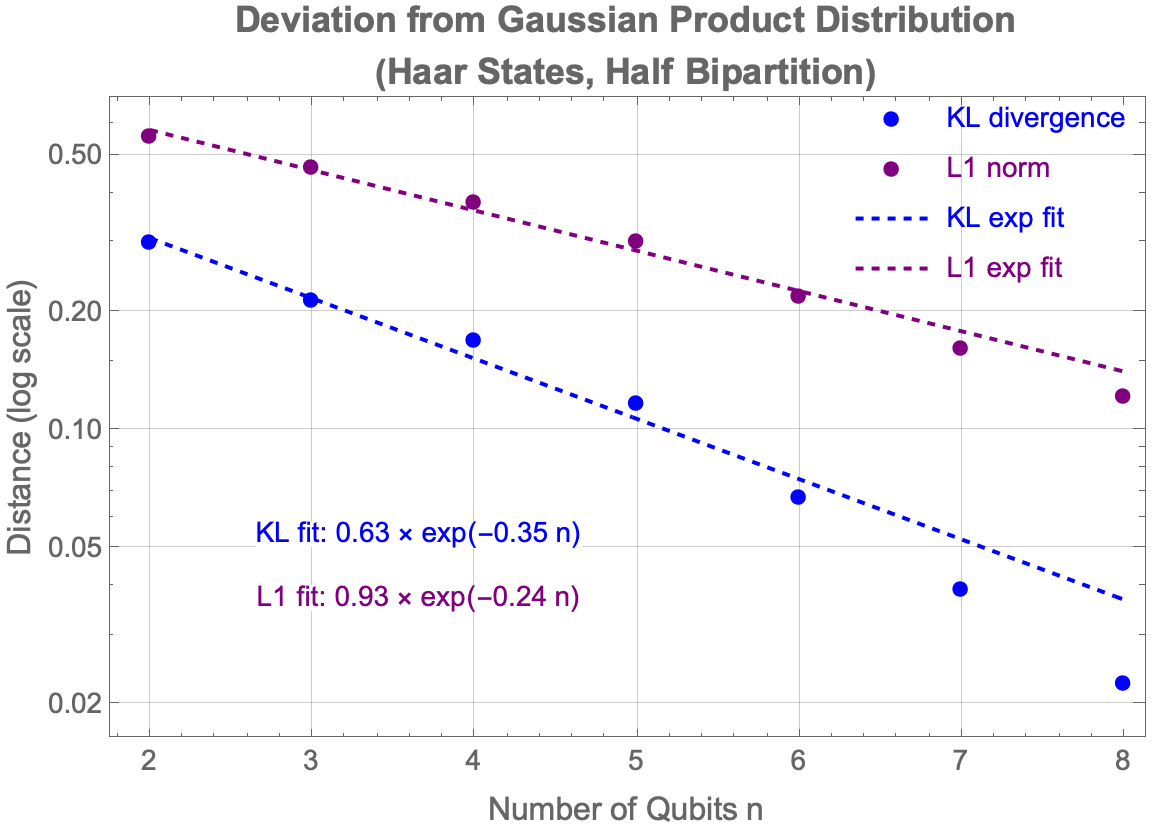}
    \caption{ {Logarithmic scale plot of Kullback–Leibler divergence $KL$ (blue) and the $L_1$ norm (violet) between the numerical PDF and the Gaussian approximation in \ref{eq:gauss_approx} for $N_{sample}=10^5$ Haar random states $\ket{\psi}$ per system size $n$.
    An exponential fit $y=a e^{-b x}$ for both $KL$ and $L_1$ is reported.}}
    \label{fig:KL_L1_decay}
\end{figure}
}

\section{Explicit calculation of the average linear SE over the orbit of factorized unitaries}\label{SE_mean_app}
The objective of this section is to show the calculations needed to compute the following average:

\begin{equation}\label{uaubW}
   I=\exv_{U_A,U_B}\Tr[Q(U_A\ot U_B \ketbra{\psi}U_A ^\dag \ot U_B ^\dag)^{\ot 4} ]\,,
\end{equation}
with $\ket{\psi}=\sum_{i=1}^{\min{d_A,d_B}} \sqrt{\lambda_i} \ket{i_A} \ket{i_B}$, with $d_A=\min(d_A,d_B)$ without loss of generality.
Here $\ket{i_A}$ and $\ket{i_B}$ are to be thought of as generic orthonormal vectors on the respective Hilbert spaces.
After a proper permutation, i.e. $\perm_{(4567)}\perm_{(345)}\perm_{(23)}$, one can write the expression using the Weingarten calculus as \cite{mele2024introduction} 
\begin{align}
    \nonumber I&=\sum_{\pi, \sigma,\gamma,\delta \in S_{4}} \sum_{i,j,k,l =1}^{d_A}\sum_{m,n,o,p=1}^{d_A} \sqrt{\lambda_i \
\lambda_j \lambda_k \lambda_l \lambda_m \lambda_n \lambda_o \lambda_p}   W_{\pi \sigma} ^A W_{\gamma
\delta} ^B \Tr[\perm^A _{\pi} Q^A ] \Tr[\perm^B_{\gamma} Q^B] \\
& \Tr[\perm_\sigma ^A \ket{i j k l}_A \bra{m n o p}] \Tr[\perm_\delta^B \ket{i j k l}_B \bra{m n o p}]\,,
\end{align}
with $W_{\pi\sigma}$ being the Weingarten function, defined as the pseudo inverse of the Gram matrix $\Omega_{\pi\sigma}:=\Tr[\perm_\pi \perm_\sigma]$.
The computation will be carried out by noting the result of the traces $\Tr[\perm_\sigma  \ket{i j k l} \bra{m n o p}]$ is symmetric between the two partitions and this means one has to compute the following terms.
The order of the permutations of $S_4$
as well as the results of $\Tr[\perm_\sigma \ket{i j k l}\bra{m n o p}]$ and $\Tr[\perm_\sigma Q]$ are summarized in Table \ref{melperm} below.
\begin{table}[h!]
    \centering
    \begin{tabular}{|c|c|c|}
    \hline
     $\sigma$    & $\Tr[\perm_\sigma \ket{i j k l}\bra{m n o p}]$ & $\Tr[\perm_\sigma Q]$  \\\hline
       Id  & $\delta_{im} \delta_{jn} \delta_{ko} \delta_{lp}$& $d^2$\\\hline
       ${(34)}$ & $\delta_{im} \delta_{jn} \delta_{ko} \delta_{lp}$ & $d$\\\hline
       ${(23)}$ & $\delta_{im} \delta_{jo} \delta_{kn} \delta_{lp}$ & $d$\\ \hline
       ${(234)}$  & $\delta_{im} \delta_{jo} \delta_{kp} \delta_{ln}$ & $1$\\ \hline
       ${(243)}$ & $\delta_{im} \delta_{jp} \delta_{kn} \delta_{lo}$ & $1$\\ \hline
       ${(24)}$ & $\delta_{im} \delta_{jp} \delta_{ko} \delta_{ln}$ & $d$\\ \hline
       ${(12)}$ & $\delta_{in} \delta_{jm} \delta_{ko} \delta_{lp}$ & $d$ \\ \hline
       ${(12)(34)}$ & $\delta_{in} \delta_{jm} \delta_{kp} \delta_{lo}$& $d^2$  \\ \hline
       ${(123)}$ & $\delta_{in} \delta_{jo} \delta_{km} \delta_{lp}$ & $1$\\ \hline
       ${(1234)}$ & $\delta_{in} \delta_{jo} \delta_{kp} \delta_{lm}$ & $d$\\ \hline
       ${(1243)}$ & $\delta_{in} \delta_{jp} \delta_{km} \delta_{lo}$ & $1$ \\ \hline
       ${(124)}$ & $\delta_{in} \delta_{jp} \delta_{ko} \delta_{lm}$ & $1$\\ \hline
       ${(132)}$ & $\delta_{io} \delta_{jm} \delta_{kn} \delta_{lp}$ & $1$ \\ \hline
       ${(1342)}$ & $\delta_{io} \delta_{jm} \delta_{kp} \delta_{ln}$ & $d$\\ \hline
       ${(13)}$ & $\delta_{io} \delta_{jn} \delta_{km} \delta_{lp}$ & $d$\\ \hline
       ${(134)}$ & $\delta_{io} \delta_{jn} \delta_{kp} \delta_{lm} $ & $1$ \\ \hline
       ${(13) (24)}$ & $\delta_{io} \delta_{jp} \delta_{km} \delta_{ln}$ & $d^2$\\ \hline
       ${(1324)}$ & $\delta_{io} \delta_{jp} \delta_{kn} \delta_{lm}$  & $d$\\ \hline
       ${(1432)}$ & $\delta_{ip} \delta_{jm} \delta_{kn} \delta_{lo} $& $d$ \\ \hline
       ${(142)}$ & $\delta_{ip} \delta_{jm} \delta_{ko} \delta_{ln}$ & $1$\\ \hline
       ${(143)}$ & $\delta_{ip} \delta_{jn} \delta_{km} \delta_{lo}$ & $1$\\ \hline
       ${(14)}$ & $\delta_{ip} \delta_{jn} \delta_{ko} \delta_{lm}$ & $d$\\ \hline
       ${(1423)}$ & $\delta_{ip} \delta_{jo} \delta_{km} \delta_{ln}$ & $d$ \\ \hline
       ${(14)(23)}$ & $\delta_{ip} \delta_{jo} \delta_{kn} \delta_{lm}$ & $d^2$\\ \hline
    \end{tabular}
    \caption{Traces of the $\perm_\sigma\ket{ijkl}\bra{mnop}$ and $Q\perm_\sigma$ operators.}
    \label{melperm}
\end{table}

After carrying out the contraction of the product of Schmidt coefficients with the proper Kronecker deltas, summing up the terms for all permutations and rearranging the terms one finally gets
\begin{equation}
        \begin{split}            
           &\exv_{U_A,U_B}[\mlin((U_A \otimes U_B \ket{\psi})]= 1-d_Ad_B\exv_{U_A,U_B}\tr[Q(U_A\ot U_B)^{\ot 4} \psi^{\ot 4} (U_A\ot U_B)^{\dag \ot 4}]=\\
            &=\alpha''+\beta''\sum_i \lambda_i ^2+\gamma''\left(\sum_i \lambda_i^2\right)^2+\delta\sum_i \lambda_i ^3+\mu\sum_i \lambda_i ^4\,,
        \end{split} 
    \end{equation} 
    with $ \delta,\mu$ defined as per Eq.~\eqref{coeff} and 
    \begin{equation}
        \begin{split}
            \alpha''&=1-\frac{1}{3} \left(\frac{8}{d_Ad_B}+\frac{2}{(d_A+3) (d_B+3)}+\frac{2}{(d_A-3) (d_B-3)}\right)\,,\\
            \beta''&=\frac{24 (d_A+d_B)}{\left(d_A^2-9\right) \left(d_B^2-9\right)}\,,\\
            \gamma''&=\frac{4 (d_A d_B+9)}{\left(d_A^2-9\right) \left(d_B^2-9\right)}+\frac{8}{d_A d_B}\,.
        \end{split}
    \end{equation}
    Proper rearrangement of the terms and collecting  $e=1- \sum_{i=1} \lambda_i^2$ yields the expression seen in Eq.~\eqref{eq:m_UAUB} which has the explicit dependence on entanglement.

\end{document}